\documentclass[floatfix,aps,prd,twocolumn,letterpaper,lengthcheck,superscriptaddress,showpacs,amssymb,amsmath,amsfonts,nofootinbib,altaffilletter,nopreprintnumbers,showpacs,longbibliography]{revtex4-1}
\usepackage[utf8]{inputenc}
\usepackage{xspace}
\usepackage{acronym}
\usepackage{booktabs}
\usepackage{todonotes}
\usepackage{graphicx}
\usepackage{amssymb}
\usepackage{color}
\usepackage{ulem}
\usepackage{wasysym}
\usepackage[colorlinks, linkcolor=blue, pdfborder={0 0 0}, breaklinks=true]{hyperref}
\definecolor{lightgray}{gray}{0.9} 
\usepackage{tabularx}
\usepackage{multirow}
\usepackage{scrextend}
\usepackage{textgreek}
\usepackage{graphicx}
\usepackage[caption=false]{subfig}
\usepackage{mathtools}
\usepackage{float}
\usepackage{soul}
\usepackage{booktabs}
\usepackage{enumitem}
\usepackage{dcolumn}
\usepackage{bm}

\usepackage{lineno} 

\DeclarePairedDelimiterX{\norm}[1]{\lVert}{\rVert}{#1}
\begin{document}

\preprint{APS/123-QED}
\title{Simulating Transient Noise Bursts in LIGO with Generative Adversarial Networks}
\author{Melissa Lopez}\email[Corresponding author: ]{m.lopez@uu.nl}
\affiliation{Institute for Gravitational and Subatomic Physics (GRASP), Department of Physics, Utrecht University,
Princetonplein 1, 3584 CC Utrecht, Netherlands}
\affiliation{Nikhef, Science Park 105, 1098 XG Amsterdam, Netherlands}

\author{Vincent Boudart}%
\affiliation{STAR Institute, Bâtiment B5, Université de Liège, Sart Tilman B4000 Liège, Belgium}

\author{Kerwin Buijsman} \thanks{This research was conducted during my studies at the University of Amsterdam.}
\affiliation{Randstad,  Diemermere 25, 1112TC Diemen, Netherlands}
\affiliation{Gravitation and Astroparticle Physics Amsterdam (GRAPPA),  Institute for Theoretical Physics Amsterdam, University of Amsterdam, the Netherlands}

\author{Amit Reza}
\affiliation{Institute for Gravitational and Subatomic Physics (GRASP), Department of Physics, Utrecht University,
Princetonplein 1, 3584 CC Utrecht, Netherlands}
\affiliation{Nikhef, Science Park 105, 1098 XG Amsterdam, Netherlands}

\author{Sarah Caudill}
\affiliation{Institute for Gravitational and Subatomic Physics (GRASP), Department of Physics, Utrecht University,
Princetonplein 1, 3584 CC Utrecht, Netherlands}
\affiliation{Nikhef, Science Park 105, 1098 XG Amsterdam, Netherlands}

\date{\today}
\begin{abstract}
The noise of gravitational-wave (GW) interferometers limits their sensitivity and impacts the data quality, hindering the detection of GW signals from astrophysical sources. For transient searches, the most problematic are transient noise artifacts, known as glitches, that happen at a rate around $ 1 \text{ min}^{-1}$, and can mimic GW signals. Because of this, there is a need for better modeling and inclusion of glitches in large-scale studies, such as stress testing the pipelines. In this proof-of concept work we employ Generative Adversarial Networks (GAN), a state-of-the-art Deep Learning algorithm inspired by Game Theory, to learn the underlying distribution of blip glitches and to generate artificial populations. We reconstruct the glitch in the time-domain, providing a smooth input that the GAN can learn. With this methodology, we can create distributions of $\sim 10^{3}$ glitches from Hanford and Livingston detectors in less than one second. Furthermore, we employ several metrics to measure the performance of our methodology and the quality of its generations. This investigation will be extended in the future to different glitch classes with the final goal of creating an open-source interface for mock data generation. 
\end{abstract}
\maketitle

\section{Introduction}
{During the first observing run (O1), the existence of gravitational-wave (GW) signal from binary black hole (BBH) coalescence was successfully proven by Advanced Laser Interferometer Gravitational-Wave Observatory (LIGO)} \cite{FirstGW}. After an upgrade of the detectors to increase their sensitivity, Advanced LIGO \citep{LIGOScientific:2014pky} started in November $2016$ the second observing run (O2), which Advanced Virgo \citep{VIRGO:2014yos} joined in August $2017$ \cite{O2_LIGO_Virgo}. Following significant upgrades, in April $2019$, the third observing run (O3) was initiated by Advanced LIGO, and Advanced Virgo \cite{O3_LIGO_Virgo}. During O1 and O2, 11 candidates were detected and 74 were detected during O3 \cite{LIGOScientific:2018mvr, LIGOScientific:2020ibl, LIGOScientific:2021djp}. In the coming years, the improvement of the second generation of interferometers and the construction of the third generation of detectors, such as Cosmic Explorer, LISA, and Einstein Telescope, will increase significantly the detection sensitivity \cite{Cosmic_Explorer, LISA, Einstein_telescope}.

While current GW search techniques for transient signals ($\lesssim$ 1 minute) have been extremely successful, their sensitivity continues to be hindered by the presence of transient bursts of non-Gaussian noise in the detectors, known as {\it glitches}. Glitches have durations typically on the order of sub-seconds, and their causes can be environmental (e.g., earthquakes, wind, anthropogenic noise) or instrumental (e.g., overflows, scattered light~\cite{Soni2020ReducingSL}), although in many cases, the cause remains unknown~\cite{Cabero:2019orq}. While much work has been done to mitigate the effect of glitches on GW searches~\cite{LIGO:2021ppb}, they remain one of the major limiting factors in the detection and parameter estimation of transient GW signals.

In this paper, we learn the underlying distribution of glitches with Machine Learning (ML) methods for better modelling an inclusion for large scales. For this aim, we employ Generative Adversarial Networks (GAN) \citep{Original_GAN} to build an artificial population of glitches and we use several metrics to test their similarity to the real input. 
This paper is structured as follows. In section \ref{sec:related_work} we introduce the current state-of-the-art of glitch identification, as well as blip glitches, which is the focus of this work. In section \ref{sec:methodology} we describe in detail the ML method employed and we give details about the data acquisition. In section \ref{sec:results} we present some examples of the generated data, we propose several statistical tests to measure the performance of our methodology and we comment on its limitations. In section \ref{sec:applications} we provide a description of several possible applications of the generated data for future investigations and in section \ref{sec:conclusion} we conclude.
\section{Gravitational-wave Detector Glitches}\label{sec:related_work}
\subsection{Identification and Classification}
Because glitches can reduce the amount of analyzable data, bias astrophysical detection, parameter estimation, and even mimic GW signals, it is fundamental to develop robust techniques to identify and characterize these sources of noise for their possible elimination. In previous LIGO and Virgo science runs, this classification was performed by visual inspection, which soon proved to be slow and inefficient \citep{Powell2015ClassificationMF}.

During O2 run, the detection rate of glitches was  $\approx 1 \text{ min}^{-1}$ ; so due to the overwhelming amount of glitches present in data. 
A promising option is to construct ML algorithms to identify and classify glitches \citep{MukherjeeSoma,Powell2015ClassificationMF,Powell2016ClassificationMF}. However, another challenge arises since a pre-labeled data set is necessary to train such algorithms. 
With this goal in mind, Zevin et al. \cite{Gravity_spy} developed pioneerwork to classify transient noise, called \textit{Gravity Spy}. In this work, both problems are addressed: volunteers provide large labeled data sets to train the ML algorithms through Zooniverse infrastructure, while ML algorithms learn to classify the rest of the glitches correctly, providing feedback to participants. In practice, a glitch time series that we wish to classify is fed to the algorithm that generates the Q-transform of its input (see \cite{Gravity_spy} for details). Then, \textit{Gravity Spy} classifier assigns a class and a confidence value $c_{GS}$ to the Q-transform of the glitch, where $c_{GS}$ represents the confidence of the label assigned.
\textit{Gravity Spy} uses a multi-class classification, and it differentiates between $23$ glitch classes and the absence of glitch inside the Q-scan in O2 \cite{Gravity_spy}.
\subsection{Blip Glitches}
\begin{figure}[!ht]
\centering
\includegraphics[width=0.5\textwidth]{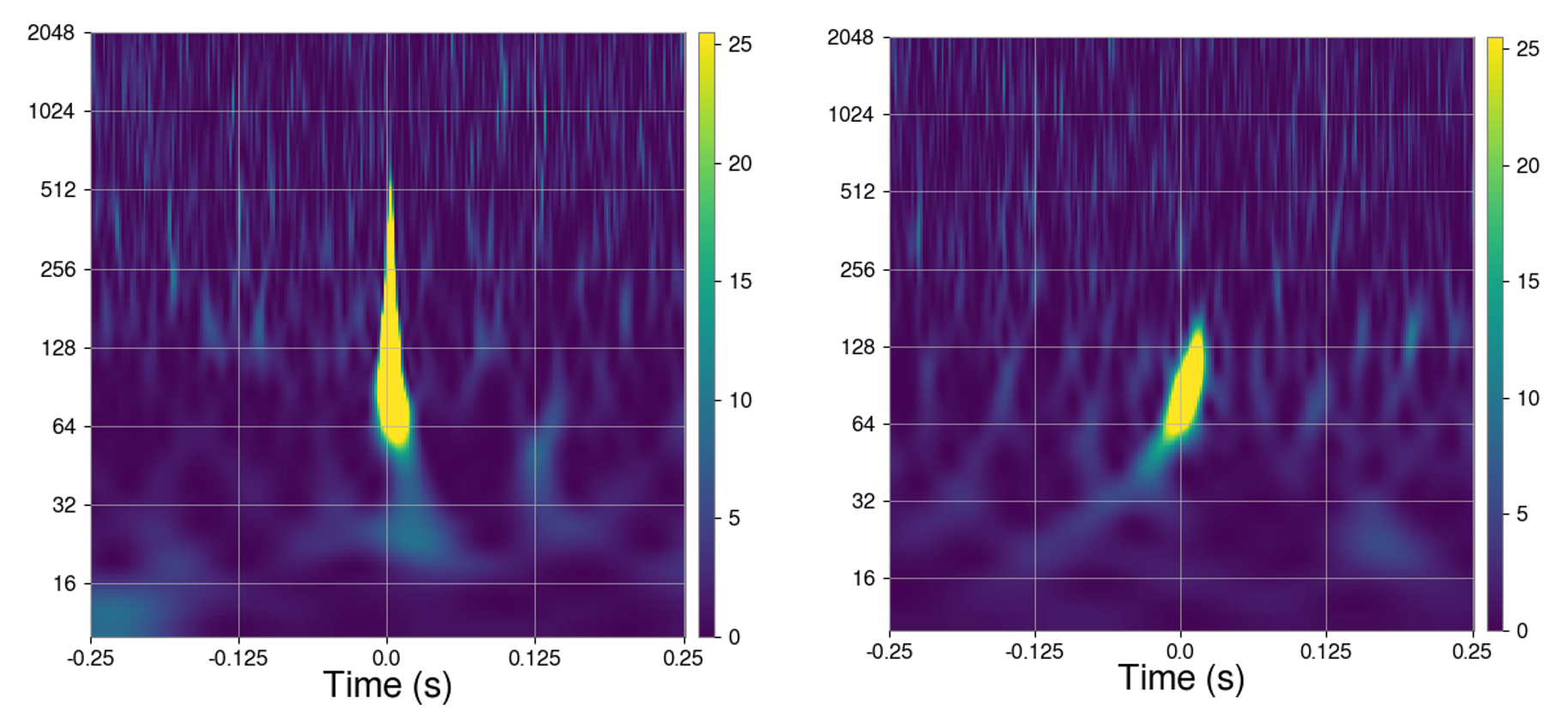}
\caption{\textit{(Left)} Q-transform of a blip glitch retrieved from \textit{Gravity Spy} \citep{Gravity_spy}. \textit{(Right)} Q-transform of an event with total mass $106.6^{+13.5}_{
-14.8} M_{\astrosun}$.}
\label{fig:glitch_IMBH}
\end{figure}

This work focuses on blip glitches due to their abundance during O2 run and their simple morphology. Blip glitches are short glitches ($\lesssim 0.2$ s)
that have a characteristic morphology of a symmetric ‘teardrop’ shape in time-frequency in the range $[30, 250]$ Hz, as we show in Fig.\ref{fig:glitch_IMBH} (left). They appear in both Livingston (L1) and Hanford (H1) detectors, which is the focus of our work, but there is also evidence of their presence in Virgo and GEO $600$ \citep{Cabero:2019orq}. 
Due to their abundance and form, blip glitches hinder both the unmodeled burst and modeled CBC searches \citep{Abbott_2018, DetChar}, with particular emphasis in compact binaries with large total mass, highly asymmetric component masses, and spins anti-aligned with the orbital angular momentum. For illustration, in Fig.\ref{fig:glitch_IMBH}, we can observe the similarities between a blip an intermediate binary black-hole chirp surrounded by O2 noise. Moreover, since there is no clear correlation to the auxiliary channels, they cannot be removed from astrophysical searches yet.
\section{Methodology}\label{sec:methodology}
ML techniques have been very successfully applied for solving a variety of tasks across different domains, and in recent times they have sparked the interest of scientists in the field of GW data analysis. A widely used ML method for pattern recognition is convolutional neural networks (CNNs), which present a grid-like topology, able to exhibit strong local spatial dependencies, allowing faster evaluation speeds \citep{Goodfellow}. 
CNNs has been successfully employed in different tasks such as identification of BBH \citep{GeorgeHuerta2018, Gabbard2018} and binary neutron stars (BNS) \cite{Menendez, Krastev2020}, detection of the early inspiral of BNS mergers \cite{Baltus, Yu2021}, supernovae identification \cite{Melissa, Chan2020, Soma2021} and glitch classification \citep{Gravity_spy, Razzano2018}, among others. See also \cite{MLreview} for an interesting review.  
CNNs can also be used to achieve pixel-wise identification of long-duration bursts in the time-frequency plane. Indeed, authors in \citep{Boudart:2022xib} built a network that learns to identify the relevant pixels in the image to later use this information to upsample  \citep{Shelhamer2017FullyCN} it into the original size. \\

ML methods are not only limited to pattern recognition tasks. GAN can learn the underlying distribution of a population to produce artificial examples from Gaussian noise. With this idea in mind, the authors in \cite{CGAN_bursts} employed a conditional GAN to burst signals, allowing them to generate multiple classes of signals with the same algorithm and to interpolate through different classes, creating mixed signals. The powerful generation capability of GAN leads the authors to foresee that it could be applied to generate artificial glitches. In the following subsection, we provide more details about GAN methodology and the architecture of our network.
\subsection{Generative Adversarial Networks}
GAN ~\cite{Original_GAN} are a class of generative algorithms in which two neural networks compete with each other to achieve realistic image generation. One network, known as the {\it generator}, is responsible for generating new images from random noise, while the other, known as the {\it discriminator}, tries to discriminate the generated images from the real training data. The generator progressively learns which features of the real images should be mimicked to fool the discriminator and save them into the latent space, which can be understood as a compressed representation of the input data learnt by the generator.
\begin{figure}[!htb]
\centering
\includegraphics[width=0.5\textwidth]{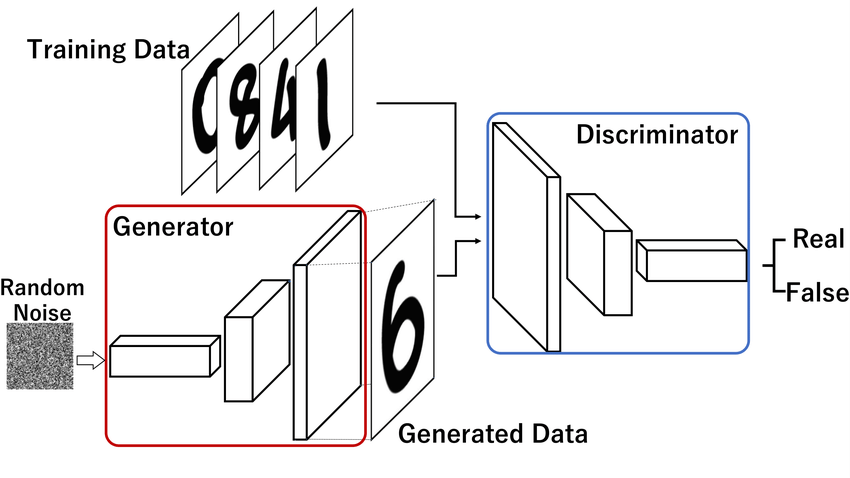}
\caption{Typical GAN architecture retrieved from \citep{Harada2019BiosignalGA}. 
}
\label{gan architecture}
\end{figure}
At the end of the training, new images are drawn by randomly taking a latent space vector and passing it to the generator, which has learned to translate it into a realistic image. 
Fig.\ref{gan architecture} shows an overview of the original architecture of GAN for generating 2D data, but all the forthcoming developments still hold for 1D data. This early approach has been shown to work well under some hyperparameter configurations \cite{Radford}. However, early GAN architecture~\citep{Original_GAN} suffers from the significant problems of vanishing gradients and meaningless loss function~\citep{Weng2019FromGT}. Wasserstein GANs~\cite{Arjovsky} (WGAN) were developed to address these issues by making use of the Earth's mover distance estimator, or Wasserstein-1 distance ($W_1$)~\citep{Kantorovich_1960}, which computes the similarities between two distributions. $W_{1}$ is evaluated through the discriminator as the training progresses and increases monotonically while never saturating, providing a meaningful loss metric even for two disjoint distributions. Since $W_{1}$  is continuous and differentiable, it yields reliable gradients, allowing us to train the discriminator till optimality to obtain high-quality generations. This change of paradigm led Arjovsky et al.\cite{Arjovsky} to reformulate the optimization problem as :
\begin{equation}
\theta_{opt} = arg\, \underset{\theta}{min}\, W_{1}(P_{x} \| P_{\Tilde{x}}) \, ,
\label{eq:1}
\end{equation}
where $W_{1}$ is evaluated between the real distribution $P_{x}$ and generated distribution $P_{\Tilde{x}}$. 
Eq.\ref{eq:1} can be written as,
\begin{equation}
\theta_{opt} = arg\, \underset{\theta}{min}\, \underset{\phi:\norm{D(x,\phi)}_{L} \leq 1}{max} L(\phi, \theta)
\label{optimization problem}
\end{equation}
with the discriminator loss:
\begin{equation}
L(\phi,\theta) = - E_{x \sim P_{x}}\big[ D(x,\phi) \big] + E_{\Tilde{x} \sim P_{\Tilde{x}}}\big[ D(\Tilde{x},\phi) \big]
\label{Loss WGAN}
\end{equation}
where $D$ and $G$ refer to the discriminator and the generator with parameters $\phi$ and $\theta$, respectively. $E_{x \sim P_{x}}$ indicates that the expression has been averaged over a batch of real samples $x$, while $E_{\Tilde{x} \sim P_{\Tilde{x}}}$ has been averaged over a batch of generated samples $\Tilde{x}$. The new condition over $\phi$ in expression Eq.\ref{optimization problem} imposes a constraint on the discriminator $D$, which must be 1-Lipschitz continuous \cite{Arjovsky}. 

In practice, this can be achieved in two ways: clipping the weights of the discriminator beyond a specific value $c$\cite{Arjovsky}, or adding a regularization term to the discriminator loss, defined in Eq.\ref{Loss WGAN}, known as gradient penalty (GP). While the first solution is a poor way to enforce the Lipschitz condition, the second solution has been widely accepted. The mathematical formulation of GP is as follows:
\begin{equation}
L_{tot} = L(\phi,\theta) + \lambda \,GP(\phi)
\label{WGANGP loss}
\end{equation}
with 
\begin{equation}
GP(\phi) = E_{\hat{x} \sim P_{\Tilde{x}}} \Big[ \big( \norm{\nabla_{x} D(\hat{x},\phi)}_2 - 1 \big)^2 \Big] \, ,
\end{equation}
where $\lambda$ is known as the regularization parameter, $\norm{\cdot}_{2}$ stands to the L2-norm and $\hat{x}$ is evaluated following:
\begin{equation}
\hat{x} = \Tilde{x}\,t + x\,(1-t) \, 
\end{equation}
with $t$ uniformly sampled $\sim [0,1]$. 
This method has shown impressive applications such as \cite{ProGAN}, but it is not restricted to WGANs  \cite{Salimans, Mescheder}. 
Nonetheless, unlike weight clipping, GP cannot enforce the Lipschitz condition everywhere, particularly at the beginning of the training. This can prevent the generator from converging to the optimal solution. 
To overcome this obstacle, Wei et al. have proposed a second penalization term to add to the loss from Eq.\ref{Loss WGAN}, called consistency term. They applied their new constraint to two perturbed versions of the real samples $x$, introducing dropout layers into the discriminator architecture. This ultimately leads to two different estimates noted $D(x')$ and $D(x'')$. The consistency term is defined as follows: 
\begin{equation}
\begin{split}
CT(\phi) & = E_{x \sim P_{x}}\big[ max \big( 0, \, d(D(x',\phi),D(x'',\phi)) \\ & + 0.1\,d(D\_(x',\phi),D\_(x'',\phi)) - M' \big) \big] \, ,
\end{split}
\label{ct_term}
\end{equation}
where d(.,.) is the L2 metric, $D\_$ stands for the second-to-last layer output of the discriminator, and $M'$ is a constant value. Wei et al. found that controlling the second-to-last layer output helps improve the performance of the WGANs. Thus, the final discriminator loss is then \cite{CTGAN}:
\begin{equation}
L_{\text{tot}} = L(\phi, \theta) + \lambda_{1}\, GP(\phi) + \lambda_{2}\,CT(\phi) \, ,
\end{equation}
with $\lambda_{2}$ being the consistency parameter. This type of WGAN was called CT-GAN, which is the one that we employ in this work. 
\subsection{Network Architecture}
The architecture of the networks has been inspired by the work presented in \cite{Radford} but nearest-neighbour (NN) sampling layers have been preferred over strided convolution layers in the generator structure. 
The convolution parameters were chosen to be fixed through the generator and discriminator layers with kernel $k=5$, no padding and stride $s=1$. $Leaky$ $ReLU(\cdot, \alpha = 0.2)$ has been chosen as the activation  layer for both discriminator and generator, with the exception of the output layer of the generator, that uses a $Tanh(\cdot)$ activation, allowing values $\sim [-1, 1]$.

\begin{figure*}[!]
\centering
\includegraphics[width=1.0\textwidth]{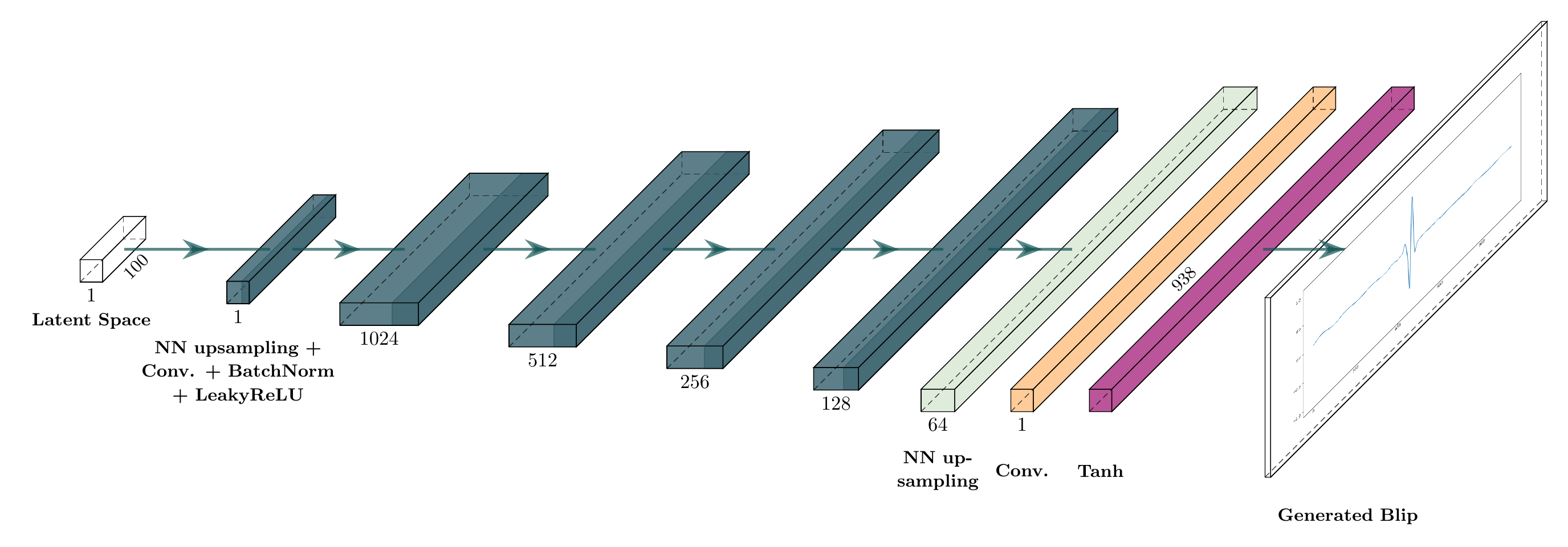}
\caption{Generator structure including NN upsampling, convolution layers and LeakyReLU activation.}
\label{generator}
\end{figure*}

\begin{figure*}[!]
\centering
\includegraphics[width=1.0\textwidth, trim={0cm 0cm 2cm 0cm}, clip]{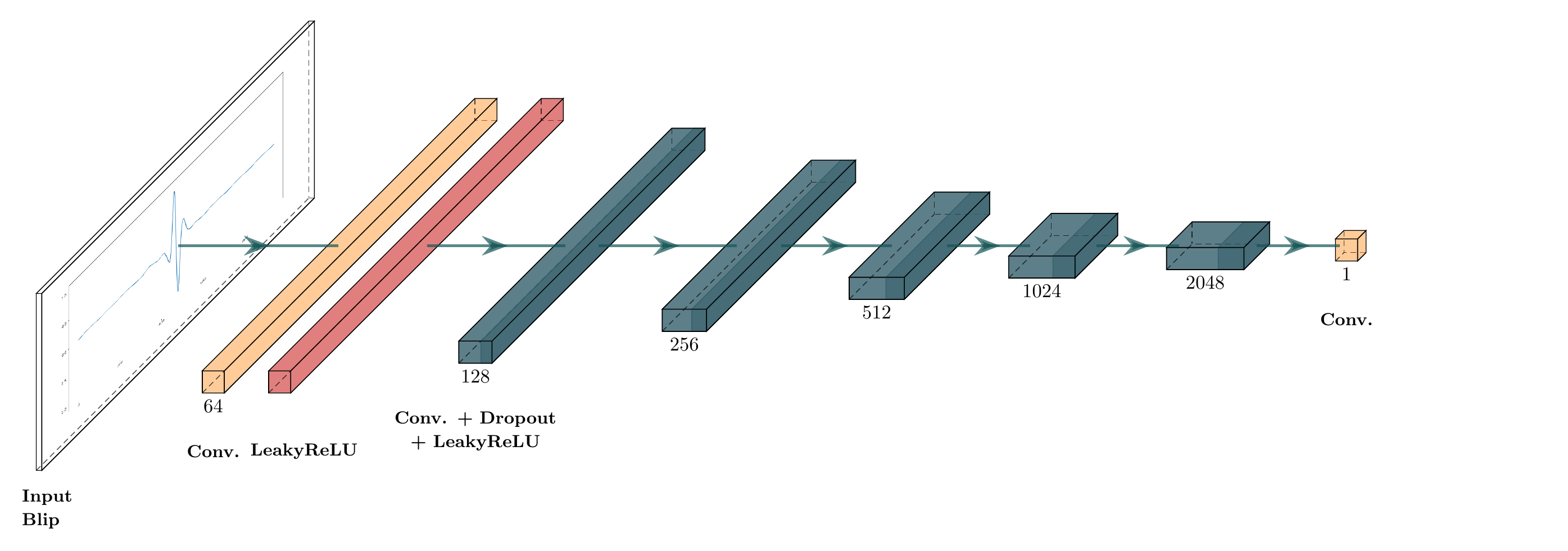}
\caption{Discriminator architecture showing strided convolutions, dropout layers and LeakyReLU activations.}
\label{discriminator}
\end{figure*}

In the generator structure (see Fig.\ref{generator}), we also employ a dilation factor of $2$, $4$, $6$, $8$ and $16$ for successive layers to enlarge its receptive field and, in turn, its expressivity power, at the exact computational cost \citep{dilation}. 
Batch normalization (BN) \citep{batchnorm} has been added to the generator architecture to make it both stable and faster to learn. The discriminator structure (see Fig.\ref{discriminator}) is composed of convolutions on which spectral normalization \cite{Gulrajani} is employed to stabilize the training. Dropout layers are added, excluding the first and last layers, which is required by the consistency term (Eq.\ref{ct_term}). 
\subsection{Training Data and Procedure}
\subsubsection{Pre-processing}
The construction of our data set strongly relies on the confidence provided by Gravity Spy. Thus, to create a high confidence data set, we select the blip glitches from L1 (Livingston) and H1 (Hanford) detectors of O2\footnote{ Data from GWOSC \url{https://www.gw-openscience.org/data/}} run that have a confidence $c_{GS}^{1} \geq 0.9 $. Glitches are surrounded by stationary and uncorrelated noise, which will hinder the learning of our machine learning method. Therefore, it is necessary to extract glitches from the stream data maintaining their original morphology. For this aim, we employ BayesLine \citep{Littenberg:2014oda} to whiten the glitches locally and BayesWave (BW) \cite{BayesWave} to extract the glitches from the uncorrelated noise. BW uses non-orthogonal continuous Morlet-Gabot wavelets to fit and reconstruct the input signal, but the selection of the model is made with a trans-dimensional Reversible Jump Markov Chain Monte Carlo  \cite{RJMCMC} that acquires a trade-off between the complexity of the model and the quality of the fit. The input signal is represented as a set of wavelets whose reconstruction is their addition. 

In our particular framework, the input provided to BW is a time series containing the blip glitch that is $2.0$ sec long. However, to avoid training the CT-GAN algorithm in irrelevant data and speed up the training phase, the samples of the final training set have $938$ data points sampled at $4096$ Hz, constituting $0.23$ sec of data.
\begin{table}[]
\caption{Size of the blip set for each detector in the different phases of the pre-processing: selection, reconstruction and evaluation.}
\label{tab:data_set}
\begin{tabular}{lcc}
\hline \multicolumn{1}{l}{Pre-processing}
& \multicolumn{1}{l}{Livingston} & \multicolumn{1}{l}{Hanford} \\ \hline
\begin{tabular}[c]{@{}l@{}} \addlinespace[0.5em] Num. blips \\ $c_{GS}^{1} \geq 0.9$\end{tabular}              & $5540$                           & $6768$                        \\
\addlinespace[0.5em]
BW output                                                                           & $5461$                           & $5612$                        \\
\addlinespace[0.5em]
\begin{tabular}[c]{@{}l@{}}Num. blips \\ $c_{GS}^{2} \geq 0.9$\end{tabular}           & $3654$                           & $3407$                        \\ 
\addlinespace[0.5em]
\begin{tabular}[c]{@{}l@{}}Num. blips \\ $c_{GS}^{2}, c_{GS}^{3} \geq 0.9$\end{tabular} & $3291$                           & $2587$                        \\ \addlinespace[0.5em] \hline
\end{tabular}
\end{table}
Since the reconstruction is not perfect, we lose around $2 \%$ and $18 \%$ of the data for L1 and H1, respectively (see Table \ref{tab:data_set}). 
To assess the quality of the reconstructions, we inject them in real whitened noise and evaluate it with \textit{Gravity Spy} classifier, selecting blips with a $c_{GS}^{2} \geq 0.9$ to generate high-quality input data. After this heavy pre-processing, the training data set is composed of around $66 \%$ and $ 50 \%$ of the initial data for L1 and H1, respectively.

\begin{figure}[!]

\includegraphics[width=1\columnwidth]{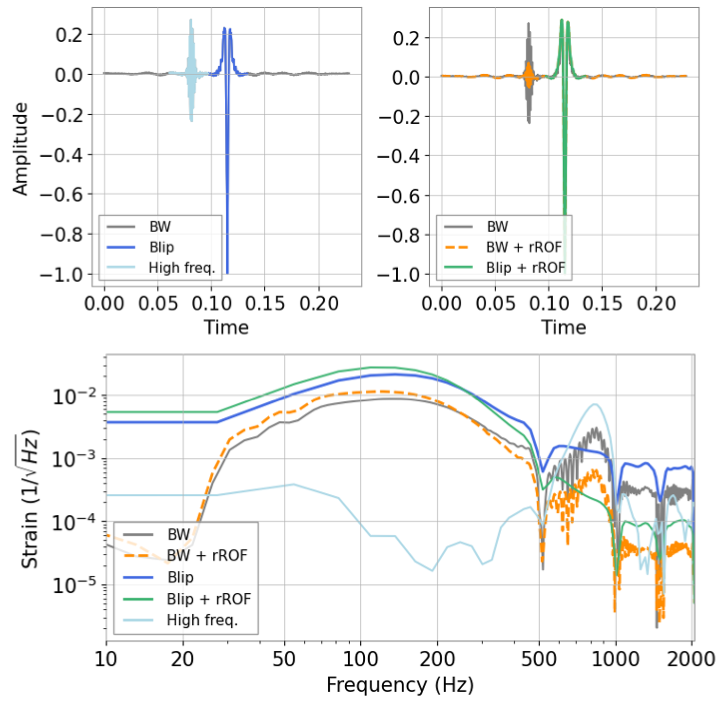}


\caption{\textit{(Top left)} Blip glitch reconstructed with BW (grey), where we colour the characteristic blip peak (blue) and the undesired high frequency contribution (light blue). \textit{(Top right)} Blip glitch reconstructed with BW (grey) and denoised with $\lambda =0.5$ (dashed orange). We colour in green the denoised characteristic blip peak. \textit{(Bottom)} Resulting amplitude spectral density (ASD) for the reconstructed blip  with BW (grey) and its denoised version with $\lambda =0.5$ (dashed orange). We also show the ASD of the characteristic peak with (blue) and without denoising (green), as well as the high frequency contribution (light blue).}
\label{fig:ex_denoised}
\end{figure}

Moreover, as it was previously mentioned, blips can be found in the frequency band $[30, 250]$ Hz, but BW might introduce certain high-frequency contributions that will hinder the learning of our machine learning algorithm. For illustration, in Fig.\ref{fig:ex_denoised} (left) we plot BW reconstruction (grey), where we coloured the characteristic blip peak (blue) and the high frequency contribution (light blue). To eliminate the high-frequency contribution, we initially set an empirical threshold to remove power excess in the surroundings of the peak. Nonetheless, some high-frequency contributions overlap with the blip and cannot be removed with this method. Thus, to minimize this contribution and generate a smoother input to enhance the learning of our model, we employ regularized Rudin-Osher-Fatemi (rROF) proposed in \cite{rROF}. 

This algorithm solves the denoising problem, $s = g + n$, where $g$ is the smooth reconstruction of glitch and $n$ is the noise, as a variational problem. The solution $g$ is computed as follows:
\begin{equation}
g_{\lambda} = \mathop{\arg \min}\limits_{g} \{\mathcal{R}(g)+\frac{\lambda}{2} \mathcal{F}(g)\} \, ,
\end{equation}
where $\mathcal{R}(g)$ is the regularization term that constrains the data, which refers to the quality of the smooth reconstruction $g$. $\mathcal{F}(g)$ is the fidelity term, which measures the $L^2$-distance between the $g$ glitch and the observed signal $s$. $\lambda$ regularises and controls the relative weight of both terms in the equation. It is important to note that his parameter needs to be tunned manually to achieve the desired level of denoising.

To assess the quality of the denoised blip glitches, we use the \textit{Gravity Spy} classifier for different $\lambda$ parameters again, and we found $\lambda = 0.5$ to be a trade-off between preserving the structure of the glitch and removing the non-smooth high-frequency contribution. In Fig.\ref{fig:ex_denoised} (right), we plot the BW reconstruction denoised with rROF (dashed orange), and the denoised characteristic blip (green). In Fig.\ref{fig:ex_denoised} (bottom), we show the amplitude spectral density (ASD) of the BW reconstruction with and without denoising (grey and dashed orange), as well as the characteristic peak with and without denoising (blue and green) and the original high-frequency contribution (light blue). We can observe that we are able to maintain the structure of the characteristic peak by damping the power of the high-frequency contribution.

\begin{figure}[ht]
\includegraphics[width =0.5\textwidth]{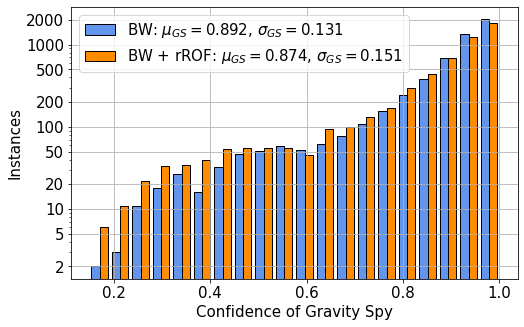}
\caption{Comparison between the reconstructed and the denoised population of blip glitches for L1. For the reconstructed set $c^2_{GS} = 0.892 \pm 0.003$ and for the denoised set $c^3_{GS} = 0.874 \pm 0.004$ at $95 \%$ confidence level.}
\label{fig:confidence_3}
\end{figure}

To verify that we are able to preserve the structure of blips according to the current state-of-the-art, we compare in Fig.\ref{fig:confidence_3} the \textit{Gravity Spy} confidence of reconstructed blips $c_{GS}^{2}$ (blue), against denoised reconstructed blips $c_{GS}^{3}$ (orange) from L1. As we can observe, both distributions are similar since they have similar means $\mu_{GS}$ and standard deviations $\sigma_{GS}$. Finally, we select the blip glitches with $c_{GS}^{2} \geq 0.9$ and $c_{GS}^{3} \geq 0.9$, to ensure the high quality of the input data of the algorithm. 
\subsubsection{CT-GAN training procedure}
During the training of the CT-GAN algorithm, both the generator and the discriminator need to be updated at similar rates to acquire stability and guarantee convergence. The task of the discriminator is more difficult since the generated samples that the discriminator intends to classify can be anywhere in the data space and change for each new iteration \cite{Kodali2017HowTT}. Hence, to assure the stability of both networks, we update the discriminator $5$ times per update of the generator, for each epoch. We employ RMSProp optimizer \citep{RMSProp} with a learning rate $ = 10^{-4}$ for both discriminator and generator, and we train the CT-GAN for $500$ epochs, where we define an epoch as the number of times the network has passed through the whole dataset. Employing GPU TITAN V with a memory of 96 Gb allowed us to use train our model in $\approx 7.75$ h.
 
\begin{figure}
\centering
\includegraphics[width=0.45\textwidth]{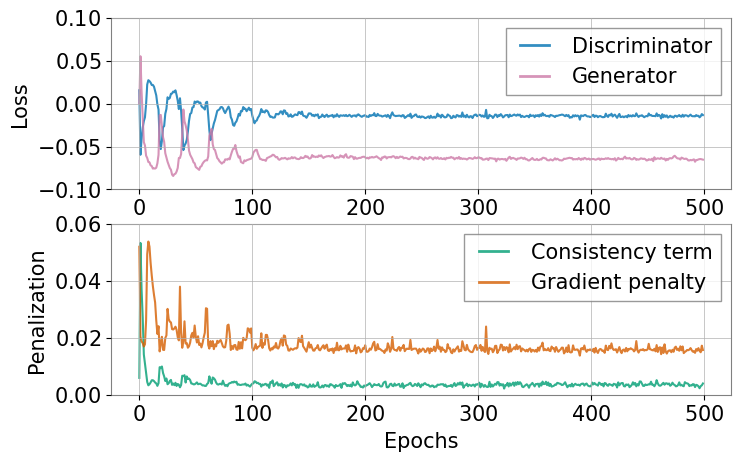}
\caption{Graph representing the discriminator loss (blue), generator loss (pink), CT (green) and GP (orange) penalisation as a function of the epochs.}
\label{fig:losses}
\end{figure}

To monitor the behaviour of the CT-GAN during the training phase, we represent the generator and the discriminator loss as a function of the epochs in Fig.\ref{fig:losses}. We can observe that both networks stabilize around epoch $100$ and continue to oscillate around values close to zero until the training is complete. After several experiments, we concluded that while CT regularised the generator, dropout regularised the discriminator and GP balanced both. This stability can also be observed in the behaviour of the CT and GP penalizations in Fig.\ref{fig:losses}, where both terms tend to zero as the network stabilizes. The values that helped the CT-GAN to converge were CT = $5$ and GP = $5$, with a dropout rate of $0.6$.

\section{Results}\label{sec:results}
\subsection{Blip Generation}
\begin{figure}[!]

{%
\includegraphics[width=1\columnwidth]{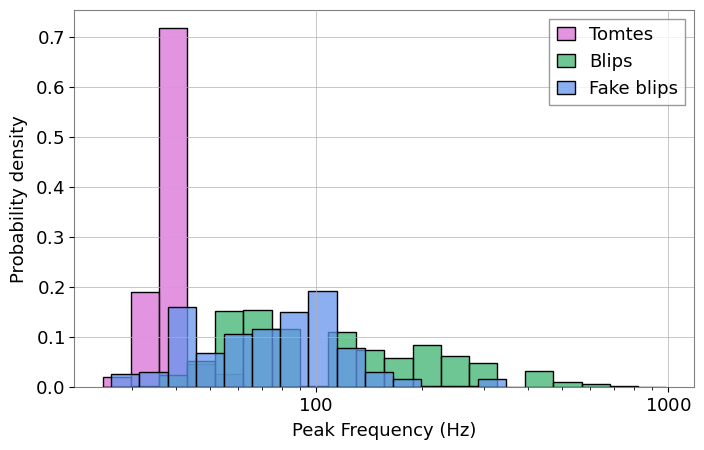}%
}\hfill

\caption{\textit{(Top)} Peak frequency for Tomte (pink) and Blip (green) from L1 retrieved from Gravity Spy \citep{Gravity_spy}, measured with Omicron spectrograms \citep{omicron}. In blue we plot the peak frequency of the artificial blips from L1.}
\label{fig:peak_duration}
\end{figure}

\begin{figure*}[]
\centering
\includegraphics[width=\textwidth]{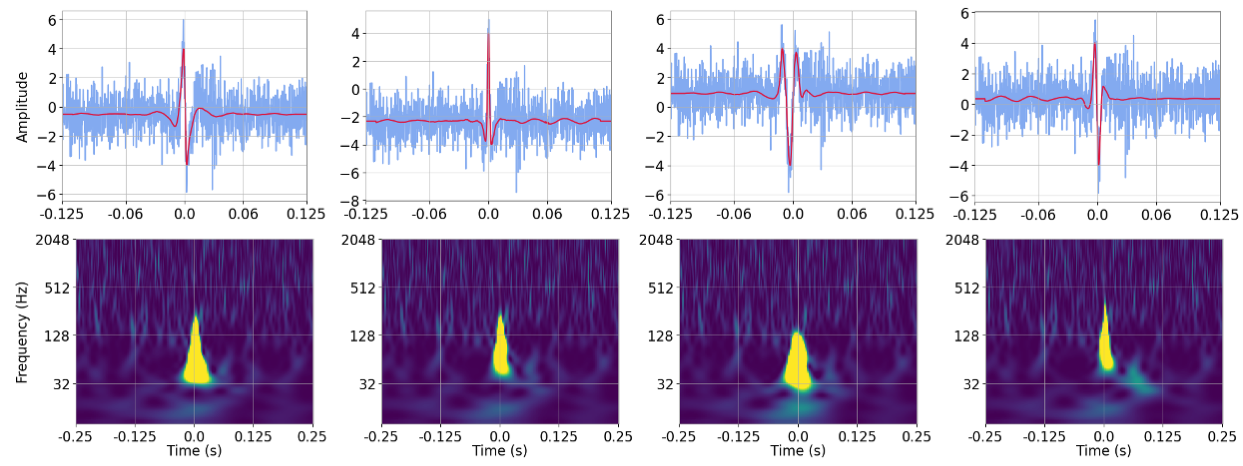}
\caption{\textit{(Top row)} Generated blip of L1 plotted as a function of time. In red we represent the rescaled whitened blip and in blue we plot its injection, both in the time domain. \textit{(Bottom row)} We show the Q-transform representation of the generated injected glitches.}
\label{fig:generated_blips}
\end{figure*}

After the training of the CT-GAN, and given a $100$-dimensional  vector drawn from a normally distributed latent space (as it common in other GAN related works), we are able to generate $10^3$ blips a blip from the input distribution of H1 and L1 in $\approx 5$ sec for both intereferometers. It is relevant to note that each blip has a length $\approx 0.23s$ with an amplitude $\in [-1,1]$, whitened and sampled at $4096$ Hz. In Fig.\ref{fig:peak_duration} we represent the peak frequency (top panel) and the duration (bottom panel) for the fake population from L1, and we compare it with Tomte and Blips. As an example, we present in Fig.\ref{fig:generated_blips}  different artificial blips from L1 in the time domain, and for visualization, we also compute their Q-transform as in \citep{Gravity_spy}. In the time-frequency representation, we can see that CT-GAN has been able to capture the distinct symmetric `teardrop' of blips in the expected frequency range $[30, 250]$ Hz. 
In Fig.\ref{fig:peak_duration}, we compare the peak frequencies of real Tomte and Blip glitches from L1 against our artificial population, where we can observe that the bulk of the distribution of fake blips is aligned with the real blip population. Furthermore, we can observe that in the time representation, we are able to reproduce different morphologies of the characteristic central peak. Even if by visual inspection it would seem that the artificial generations are closely related to the real blips from O2, it is necessary to perform a statistical test to assess the performance of CT-GAN. 

\subsection{Assessing Performance}\label{sec:metrics_definition}
We employ four different methods to assess the quality of the population. On the one hand, we employ their Q-scan representation to evaluate our artificial population with the current state-of-the-art. On the other hand, we analyze their morphology in the time domain to take into account the phase information: 

\begin{itemize}
\item \textbf{Gravity Spy classifier:} We can inject the generated glitches in real whitened noise from O2 (see Fig.\ref{fig:generated_blips}) and evaluate them with \textit{Gravity Spy}, which will return a confidence value $c_{GS}$ and a class label. We use the same noise strain for each generated glitch to provide the classifier with a fair comparison. Since the generated blip has an amplitude $\in [-1,1]$, we can re-scale it according to a desired optimal signal-to-noise ratio ($\rho_{opt}$). For this aim, we relate $\rho_{opt}$ to the scaling parameter $\alpha$ by modifying Eq 4.3 from \cite{definition_SNR} as:
\begin{equation}
\rho_{opt} = 4 \alpha \int^{f_{max}}_{f_{min}} \frac{|\Tilde{g}(f)|^2}{S_n(f)} df \, ,
\label{eq:scaling_snr}
\end{equation}
where $\Tilde{g}(f)$ represents the artificial blip and $S_{n}$ is the power spectral density (PSD) of the fixed real whitened noise. One of the main drawbacks of this method is that it is computationally intensive ($\approx 90$ s/glitch) because it is necessary to calculate the Q-transform of the input time series.
\item \textbf{Wasserstein distance $(W_{1})$}: As explained in subsection \textit{ii}, the Wasserstein distance is continuous and never saturating, allowing us to keep track of the quality of the generated samples during the training. For further mathematical details, a formal definition can be found in \cite{Arjovsky}. This metric is then an adequate tool to compare real and generated glitches. This method is fast and efficient since the computation is performed in the time domain ($\approx 0.0026$ s/glitch). 
\item \textbf{Match function $(M_f)$:} To compute the similarity between two signals, we can also use the match function, which returns the match between both signals \cite{pycbc}. The match can be defined as the inner product between two normalized signals maximized over time ($t$) and phase ($\phi$) \cite{definition_match},
\begin{equation}
M_{f}(a, b) := \max_{t, \phi}{\langle\hat{a},\hat{b}\rangle}.
\end{equation}
Since the signals are noise-free, we do not employ any PSD for normalization. This calculation is performed in the frequency domain, and it is also fast and efficient ($\approx 0.0032$ s/glitch).

\item \textbf{Normalized cross-covariance ($k(X, Y)$):} Assuming two random processes $X$ and $Y$, their cross-covariance between time $t_{1}$ and $t_{2}$ is defined as:
\begin{equation}
K_{X,Y}(t_{1}, t_{2}) \equiv E[(X_{t_{1}}-\mu(X_{t_{1}})) \overline{(Y_{t_{1}}-\mu(Y_{t_{1}}}))]
\end{equation}
To obtain the normalized cross-covariance coefficient, we divide the cross-covariance over the standard deviation of each random process. The maximum value of this magnitude is the metric employed to measure the similarity between two signals, as defined below:
\begin{equation}
k = \max{\big(\frac{K_{X,Y}(t_{1}, t_{2})}{\sigma_{X} \sigma_{Y}}\big)}
\end{equation}
This calculation, which is also in time domain, is most efficient ($\approx 0.0011$ sec/glitch).
\end{itemize}
\subsubsection{Gravity Spy}
For this procedure, we inject each generated blip in real whitened detector noise and re-scale it according to Eq.\ref{eq:scaling_snr} to fix $\rho_{opt}$. We can compute the confidence of \textit{Gravity Spy} as a function of the optimal SNR $\rho_{opt} \in [0.1, 18.2]$. This process is conducted on $10^{3}$ blip glitches of each detector population.

\begin{figure}[ht]
\centering
\includegraphics[width=0.5\textwidth]{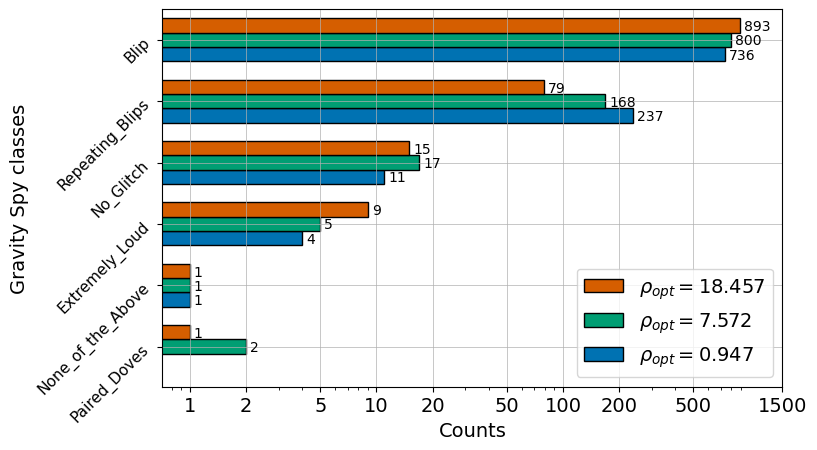}
\caption{Histogram of predicted \textit{Gravity Spy} classes for $10^3$ generated blips from H1.}
\label{fig:class_gs}
\end{figure}
In Fig.\ref{fig:class_gs}, we plot the classification labels, with maximum classification probability, for different $\rho_{opt}$ of H1 population, while we present the results of L1 in  Appendix \ref{sec:appendix}. We can observe that the dominant class is  $Blip$ and that the number of glitches in this class increments by increasing $\rho_{opt}$, in opposition to other classes. Interestingly, when $\rho_{opt} = 0.947$, meaning that the artificial blip is not visible by eye in the Q-transform, around $500$ artificial glitches are labeled as $Blip$.

One could think that this type of behaviour would be expected since CNNs are able to ``see" patterns that are invisible to the human eye, but the classifier is able to recognize glitches up to a certain threshold (Omicron SNR $\geq  7.5$ \cite{omicron}). Another reason might be that the training set of \textit{Gravity Spy} is imbalanced, so the classifier is biased towards the larger classes such as $Blip$. Hence, it seems that \textit{Gravity Spy} has a certain degree of miss-classification, so we employ other metrics to test the performance of our CT-GAN.
\subsubsection{Wasserstein distance,  match function and normalised cross-covariance during testing}
\label{sec:results_metrics}
To measure the performance of the network, we use some alternative methods, namely Wasserstein distance ($W_{1}$), match function ($M_{f}$), and normalized cross-covariance ($k$). These metrics are employed to calculate the similarity between two different artificial blips $b_{1}$ and $b_{2}$, but we can also use them to calculate the similarity between a single artificial blip $b_{F}$ and the real population ($B_{R}$) or the artificial population ($B_{F}$) from each detector. Such procedure is as follows:
\begin{enumerate}
\item We use a certain similarity distance $m$ to measure the distance between blip $b_{j}$ and a population $B$.
\item For each blip $b_{i} \in B$ we compute $m_{j, i}(b_{j}, b_{i})$, which yields a set of measurements $M_{j}$.
\item We obtain the mean and the standard error of the previous set as $\mu(M_{j}) \pm \epsilon(M_{j})$ at $99.7 \%$ confidence interval.
\end{enumerate}
The latter is the measure of similarity between the population $B$ and $b_{j}$. Note that the numerical meaning of Wasserstein distance, match function, and normalized cross-covariance are different. For the previous example,
\begin{itemize}
\item If $b_{j}$ is a reliable generation then $W_{1}(B, b_{j}) \approx 0$, while $M_{f}(B, b_{j}) \approx 1$ and $k(B, b_{j}) \approx 1$.
\item If $b_{j}$ is an anomalous generation then $W_{1}(B, b_{j}) \gg 0$, while $M_{f}(B, b_{j}) \ll 1$ and $k(B, b_{j}) \ll 1$.
\end{itemize}
Since we are dealing with real data, the real population $B_{R}$ contains not only blips but also certain misclassifications. If the CT-GAN had learned the underlying distribution of the data, we would expect that the real population $B_{R}$ and the artificial population $B_{F}$ had a similar distribution, where reliable generations would be located in the bulk of the distribution. In contrast, anomalous blips would be located in the tails. Hence, under this assumption, we would expect that, given a metric $m$, the similarity distance between the real and artificial distribution $m(B_{R}, B_{F})$, should be linearly related to the similarity distance of the artificial distribution against itself, $m(B_{R}, B_{F})$.
\begin{figure}[!]
\subfloat[\label{fig:metrics_real_fake_wass}{Similarity distance: Wasserstein distance ($W_1$)}]{%
\includegraphics[width=1\columnwidth]{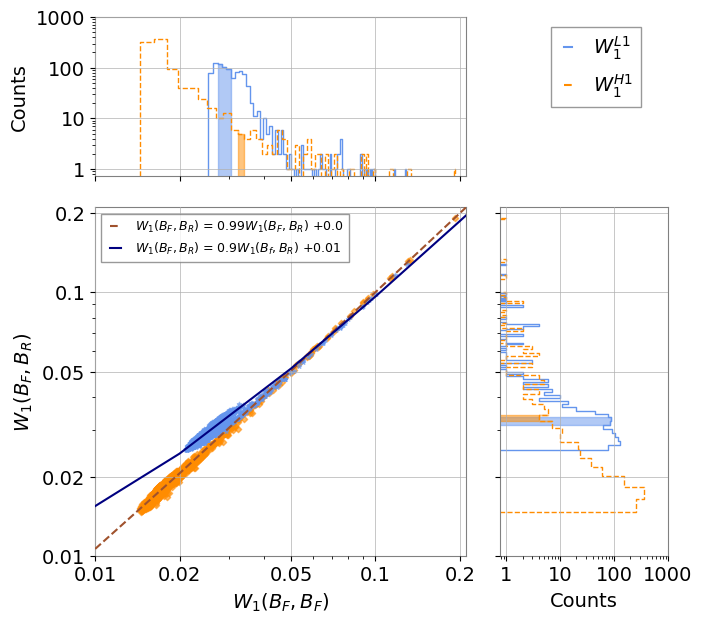}%
}\hfill
\subfloat[\label{fig:metrics_real_fake_match}{Similarity distance: match function ($M_f$) and normalised cross-covariance ($k$)}]{%
\includegraphics[width=1\columnwidth]{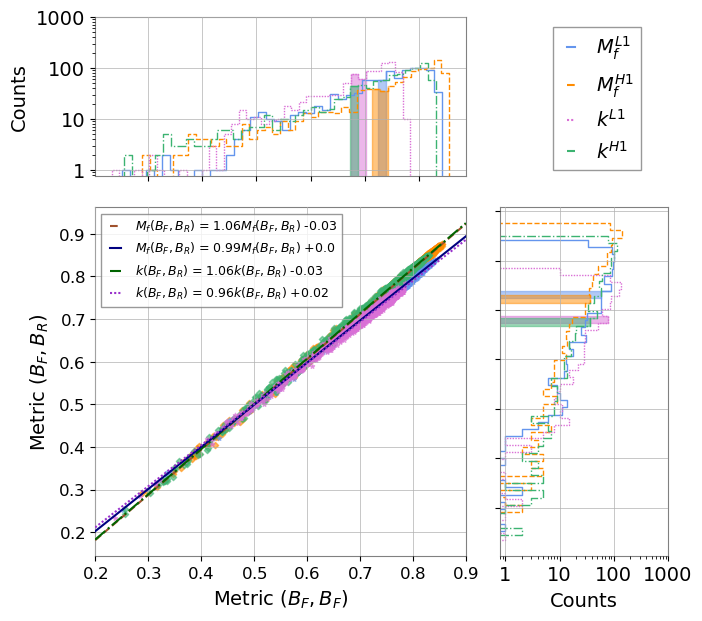}%
}\hfill
\caption{\textit{(Top)} We represent the joint and marginal distributions of $W_1(B_{R}, B_{F})$ and $W_1(B_{F}, B_{F})$ for L1 (blue) and H1 (orange) and their best fit. \textit{(Bottom)} We represent the joint and marginal distributions of the pairs $[M_f(B_{R}, B_{F}), M_f(B_{F}, B_{F})]$ and  $[k(B_{R}, B_{F}), k(B_{F}, B_{F})]$ for L1 (blue and pink) and H1 (orange and green), as well as their best fit. The coloured regions in the marginal distributions represent the confidence interval at $6$ standard deviations.}
\label{fig:metrics_real_fake}
\end{figure}
\begin{figure*}[!]
\subfloat[{Wasserstein distance ($W_{1}$)}]{%
\includegraphics[width=0.65\columnwidth]{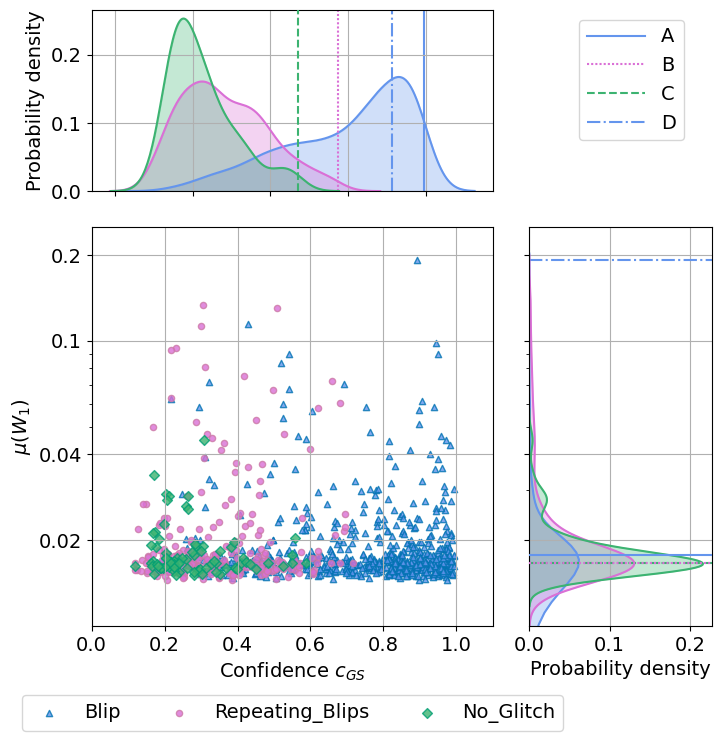}%
}\hfill
\subfloat[{Match fucntion ($M_{f}$)}]{%
\includegraphics[width=0.65\columnwidth]{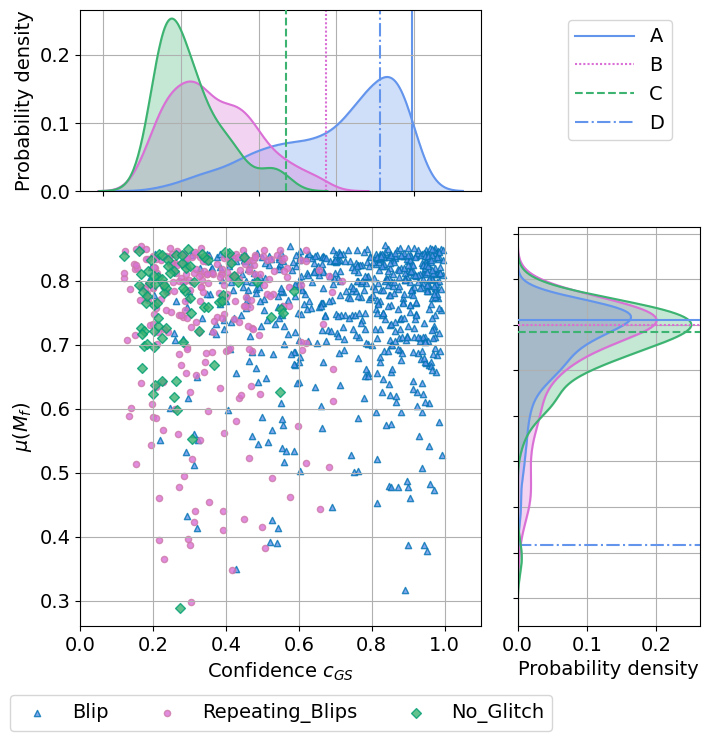}%
}\hfill
\subfloat[{Normalised cross-covariance ($k$)}]{%
\includegraphics[width=0.67\columnwidth]{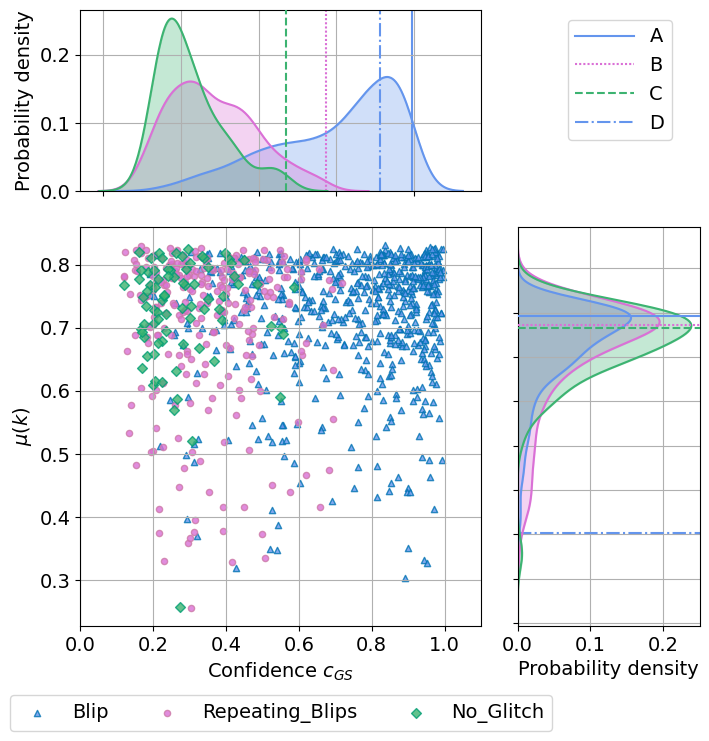}%
}
\caption{Joint and marginal distribution of \textit{Gravity Spy} confidence $c_{GS}$ at $\rho_{opt}=18.46$ against different metrics for different glitch classes for H1: $Blip$, $Repeating\_Blips$ and $No\_Glitch$. We mark in the marginal distributions selected generated glitches A (solid blue), B (dotted pink), C (dashed green) and D (dash dotted blue).}
\label{fig:gs_metrics}
\end{figure*}
In Fig.\ref{fig:metrics_real_fake}, we plot the joint and marginal distribution of both comparisons for different similarity distances: Wasserstein distance ($W_{1}$), match function ($M_f$) and normalised cross-covariance ($k$), and present the results from the least-squares estimate for each detector. Furthermore, in Table \ref{tab:data_set} we present the Pearson coefficient resulting from the least-squares estimate, which represents the linear correlation between both variables \citep{Probability_and_Stats|Pearson}.  

\begin{table}[h]
\centering
\caption{Pearson coefficient for different metrics and detectors.}
\label{table:pearson_coefficients}
\begin{tabular}{lcc}
\hline
& \multicolumn{1}{l}{Livingston} & \multicolumn{1}{l}{Hanford} \\ \hline
\begin{tabular}[c]{@{}l@{}} \addlinespace[0.5em] Wasserstein\\ distance\end{tabular}        & $0.993\%$                        & $0.999\%$   \                  \\
\addlinespace[0.5em]
Match function                                                        & $0.999\%$                        & $0.999\%$                    \\
\addlinespace[0.5em]
\begin{tabular}[c]{@{}l@{}}Normalised\\ cross-covariance\end{tabular} & $0.996\%$                       & $0.999\%$                     \\ \hline
\end{tabular}
\end{table}

We observe that the resulting slopes (Fig.\ref{fig:metrics_real_fake}) and the Pearson coefficients (Table \ref{tab:data_set}) for each metric and each detector are close to $1.0$, meaning that both variables have a very strong linear relationship and compatibility. Thus, all similarity distances indicate that the bulk of the population is constituted by reliable blips, with the presence of some anomalous generations that can be identified by fixing an empirical threshold. Therefore, since the generated blips represent the artificial and real populations, we conclude that the CT-GAN has learned the underlying distribution of blips from L1 and H1.
\subsection{Assessing Poor Generations}
When dealing with real data, one must bear in mind that certain anomalies might be present in the data. In our particular context, our data sets might contain glitches that have a distinct morphology from the mean of the population. Such differences might not be visible in a Q-transform representation, so \textit{Gravity Spy} might introduce certain miss-classifications that contaminate the input dataset. Since CT-GAN is able to learn the underlying distribution, it can also generate non-blip glitches that are in the tails of the distribution. For certain studies, the presence of anomalies might be counterproductive, so differentiating reliable from anomalous generations is crucial. For this aim, we propose several metrics to identify these miss-generations.

To use {Gravity Spy} classifier, we inject the generated blips in real whitened noise with a fixed optimal SNR $\rho_{opt} = 18.46$, according to Eq.\ref{eq:scaling_snr}. From the classification, we select the generated blips that belong to the three dominant classes: $Blip$, $Repeating\_Blips$, and $No\_Glitch$. 

\begin{figure*}[htb]
\subfloat[{Glitch A}]{%
\includegraphics[width=0.53\columnwidth]{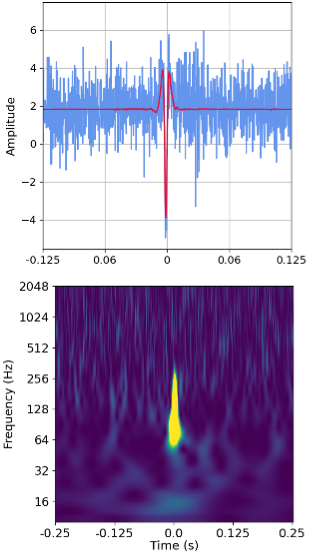}%
}\hfill
\subfloat[{Glitch B}]{%
\includegraphics[width=0.51\columnwidth]{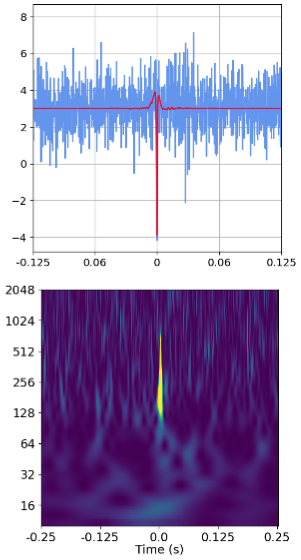}%
}
\subfloat[{Glitch C}]{%
\includegraphics[width=0.5\columnwidth]{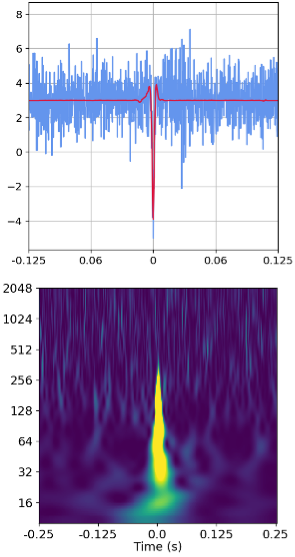}%
}\hfill
\subfloat[{Glitch D}]{%
\includegraphics[width=0.49\columnwidth]{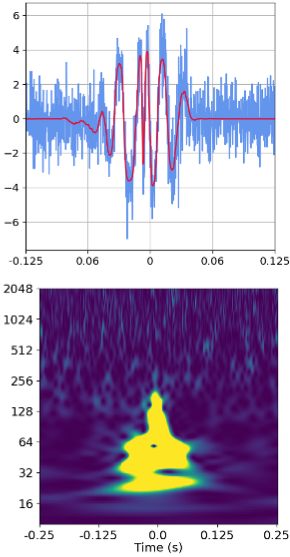}%
}
\caption{Time series representation (top row) and Q-scan representation of selected glitches from H1}
\label{fig:generations}
\end{figure*}

In Fig.\ref{fig:gs_metrics} we plot the joint and marginal distribution as probability densities of {Gravity Spy} confidence against the alternative metrics for H1  {(see Appendix \ref{sec:appendix} for details about L1)}. We can observe that according to \textit{Gravity Spy} $Blip$, $Repeating\_Blips$, and $No\_Glitch$ seem to belong to distinct probability densities. However, according to the alternative metrics, the probability densities remain centered according to a certain value for different classes. Furthermore, there seems to be no correlation  between {Gravity Spy} confidence and other metrics in the joint distribution, so to further understand our results, we proceed to inspect the results by  selecting examples,
\begin{itemize}
\item \textbf{Glitch A:} This glitch is labeled as a $Blips$ with a high confidence according to \textit{Gravity Spy} ($c_{GS} \approx 0.99$). Furthermore, the chosen metric has situated this glitch in the bulk of the distribution, meaning that it is a reliable blip generation.
\item \textbf{Glitch B:} This glitch is labeled as a $Repeating\_Blips$ with a confidence $c_{GS} \approx 0.72$. However, according to our metrics, it is a reliable generation.
\item \textbf{Glitch C:} This glitch is labeled as a $No\_Glitch$ with a confidence  $c_{GS} \approx 0.59$. However, according to our metrics, it is also a reliable generation.
\item \textbf{Glitch D:} This glitch is labeled as a $Blips$ with a high confidence according to \textit{Gravity Spy} ($c_{GS} \approx 0.89$). Nonetheless, the chosen metric has situated this glitch in the tail of the distribution, meaning that it is an anomalous blip generation.
\end{itemize}
In Fig.\ref{fig:gs_metrics}, we can observe that according to the alternative metrics, glitches A, B and C are situated around the center of the probability density, while glitch D is located in the tails. Moreover, for further visualization in Fig.\ref{fig:generations}, we present the selected in the time domain, and we also plot their Q-transforms. We can observe that while glitches A, B, and C seem to have a similar shape and magnitude, they differ from anomalous glitch D. Moreover, with these metrics, we are able to identify anomalous generations that deceive {Gravity Spy} classifier, and their exclusion from the generated data set can be performed by imposing a threshold.

\subsection{Limitations}

The main shortcoming that we encountered when training the CT-GAN was the limited amount of data preserved after the heavy pre-processing. CT-GAN needs a large amount of samples to learn the underlying distribution, which might be a limitation when extending our methodology to other classes of glitches that are less common in the LIGO/Virgo streams. Nonetheless, some researchers are developing techniques to tackle this limitation that we will explore in future works \citep{Karras2020TrainingGA}.

Another relevant shortcoming of this study is the fact that the quality of our input data set strongly relies on BW reconstruction and \textit{Gravity Spy} classification. In our particular case, blip glitches have a simple morphology, but some undesired contributions were introduced by BW, and some miss-classifications were introduced by Gravity Spy. Other glitches might be even harder to extract and/or classify with the current state-of-the-art due to their complex form, which in turn will hinder the performance of our CT-GAN. Moreover, longer and more complex glitches will need better architectures to be able to learn the underlying distribution of the data.

\section{Applications}\label{sec:applications}
In the following we provide examples of possible applications that can be explored in future works:
\begin{enumerate}[label=\itshape\Alph*)]
\item \textit{Glitch population statistics:} Learning the distributions of glitches allows us to understand their populations further and compare their different characteristics. In this way, we can develop statistics to analyze their morphologies, populations, and production rates in more detail as it was discussed in \citep{2021arXiv211002689A}. For illustration employing generated blips, we have reduced the dimensionality of the artificial population of L1 with Principal Component Analysis (PCA) \citep{PCA}. By visual inspection, we can see three main clusters that we classify with Gaussian Mixture \footnote{For both algorithms, we employ Scikit-learn implementation \citep{scikit-learn}} \citep{Reynolds2009}. Each point represents a single fake blip in PCA space coloured according to their cluster label. 
\begin{figure}[h]
\centering
\includegraphics[width=1\columnwidth]{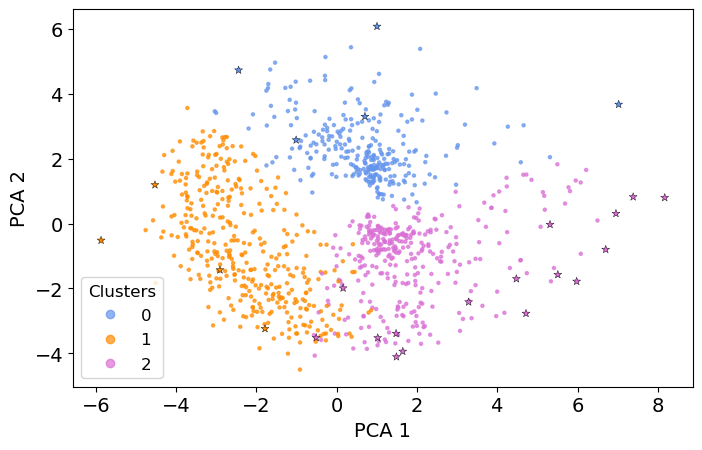}
\caption{PCA representation of fake blip population of H1, clustered with Gaussian Mixture. The $5 \%$ most anomalous blips according to the distance $W_1(B_F, B_F)$ are marked with a star.}
\label{fig:example_metric}
\end{figure}
Furthermore, we have marked with a star $5 \%$ of the most anomalous blips present in the population, according to their distance $W_1(B_F, B_F)$. It would be interesting to investigate the differences between the clusters in these distributions in future work. Another possibility would be to link the features of the blip glitches with their representation in the latent space of the CT-GAN, as it was proposed in \citep{Shen2020InterpretingTL}.  
\item \textit{Glitch template banks:} It is well-known that blip glitches have a similar morphology to intermediate black holes (IMBH), which hinders the detection of such events. With our generator, we could create glitch templates to use matched-filtering techniques in unknown signals to compute a ranking statistic and weight it in the likelihood function of detection pipelines. In this way, we would provide another metric to distinguish blip glitches from IMBH. We could use the standard matched-filter method \cite{allen2012findchirp} (See Eq.\ref{Eq:SNR}) to compute the SNR time-series for a specific glitch template. However, performing a matched filtering operation for a large glitch bank will be a huge task as computational time will increase drastically. We need to handle the scalability issue of the computational time of performing matched filtering with the increased number of glitch templates as we would expect to manage many glitch templates. We can resolve this scalability issue if we adapt the matched filtering framework used in the GstLAL \cite{cannon2021gstlal, hanna2020fast} pipeline for the searches of GW signals from CBC sources. We observed that a few numbers of basis obtained using Singular Value Decomposition (SVD) \cite{van1996matrix, cannon2010singular, kulkarni2019random, reza2021random} can also represent the glitch templates, and those basis can be used to get the matched filter output quickly. The computational time-complexity of matched filtering can be reduced as the required number of basis vectors is much less than the number of glitch templates. 
To show the efficacy of this framework, we generated $10^{3}$ glitches for the L1 detector using our proposed CT-GAN-based glitch generator. We used $1$ second data, sampled at $4096$ Hz for this study. The data contains an injected glitch and colored Gaussian noise with aLIGO Zero Detuned High Power (ZDHP) noise power spectral density \cite{shoemaker2010advanced}. Since the generated glitches are around $0.23$ second ($938$ data points) sampled at $4096$ Hz, we padded them with zero and made them $1$ sec long to generate the noisy data. The amplitudes of the injected glitch were adjusted for a target Signal to Noise Ratio (SNR) of $10$. Further, we used ZDHP to whiten the data and the glitch templates. We computed the SNR time-series for each glitch template based on (a) standard matched-filter method \cite{allen2012findchirp} as follows: 
\begin{equation}
\langle s(t), g(t) \rangle = 4 \, \text{Re} \, \int_{0}^{\infty} {\frac{\tilde{s}(f) \, \tilde{g}^{*}(f)}{S_{n}(f) } \, df } \, ,
\label{Eq:SNR}
\end{equation}
where the term $S_{n}(f)$ is defined as the one-sided power spectral density (PSD). The square root of Eq.\ref{Eq:SNR} is \textcolor{black}{termed} as SNR. \\
(b) SVD based matched filter \cite{cannon2010singular} in which a set of few top basis vectors have been computed from glitch-matrix first. Since each glitch template has $4096$ data points, therefore the dimension of the glitch matrix is of size $10^{3} \times 4096$ after stacking all the glitches together. After that, the basis vectors are matched filter against data, and the SNR time-series has been computed by combining coefficients of each glitch and matched filter output obtained based on basis and data. For our example, we obtained that $10$ top-basis vectors are sufficient to represent those $10^{3}$ glitches, as it can be observed in Fig. \ref{fig:sing-values}. It shows that the singular values of a set of $10^{3}$ glitches are fall steeply, which implies a few top-basis (e.g., $10, 20$) can be used to represent those glitches. We have chosen the number of top-basis ($\ell$) $= 1, 5, 10$ and reconstructed the glitches in our analysis. We have computed the reconstruction error for each glitch as follows:
\begin{equation}
\epsilon_{\alpha} = \frac{\|g_{\alpha} - \hat{g}_{\alpha} \|_{2}}{{\|g_{\alpha}}\|_{2}} \, ; \alpha = 1, 2, \cdots, 10^{3}\,  
\label{Eq.Glitch_Reconstruction_Error}
\end{equation}
where $\hat{g}_{\alpha}$ is the reconstructed whitened glitch based on $\ell = 1, 5, 10$ basis vectors respectively and $\| \|_{2}$ represents $\text{L}_{2}$ norm, and $\alpha$ is the number of total glitch templates. 
We also computed the fractional SNR-loss \cite{cannon2010singular} for each glitch templates based on following definition: 

\begin{equation}
\frac{\delta \rho_{\alpha}}{\rho_{\alpha}} = \frac{|\rho_{\alpha}| - |\hat{\rho}_{\alpha}|} {|\rho_{\alpha}|} \, ; \alpha = 1, 2, \cdots, 10^{3}
\label{Eq.SNR_Error}
\end{equation}
With the increasing number of basis, the relative reconstruction error should be decreased. To establish this statement, in Fig.\ref{fig:Glitch-Reconstruct-Error}, we choose three different cases with varying $\ell = 1, 5, 10$. Fig.\ref{fig:Glitch-Reconstruct-Error} shows the probability density of the relative error $\epsilon_{\alpha}$ for $\ell = 1, 5, 10$ respectively. The figure shows that relative error is less for $\ell = 10$, whereas the relative error is high for $\ell = 1$. Similarly, we obtained the fraction SNR loss for all glitch templates for these three cases. Fig. \ref{fig:SNR-Reconstruct-Error} shows the construction of glitch and SNR time-series based on $\ell = 1, 5, 10$ number of basis respectively. 
Both plots show that $\ell = 10$ is sufficient to reconstruct the whitened glitches and represent the SNR time series. If we increase the number of basis, the reconstruction errors ($\frac{\delta \rho_{\alpha}}{\rho_{\alpha}}$, $\epsilon_{\alpha}$) can be reduced but matched filtering cost would increase. Hence, we need to choose a minimal set of the basis for which computation cost and also the reconstruction errors are low. We have chosen $l = 10$ as that minimal number for this specific example. 

In a follow-up work, we will explore the possibility the construction of a  glitch bank construction, with a discussion on how to obtain ranking statistics, and signal consistency tests.

\begin{figure}[!]
\includegraphics[width=0.9\columnwidth]{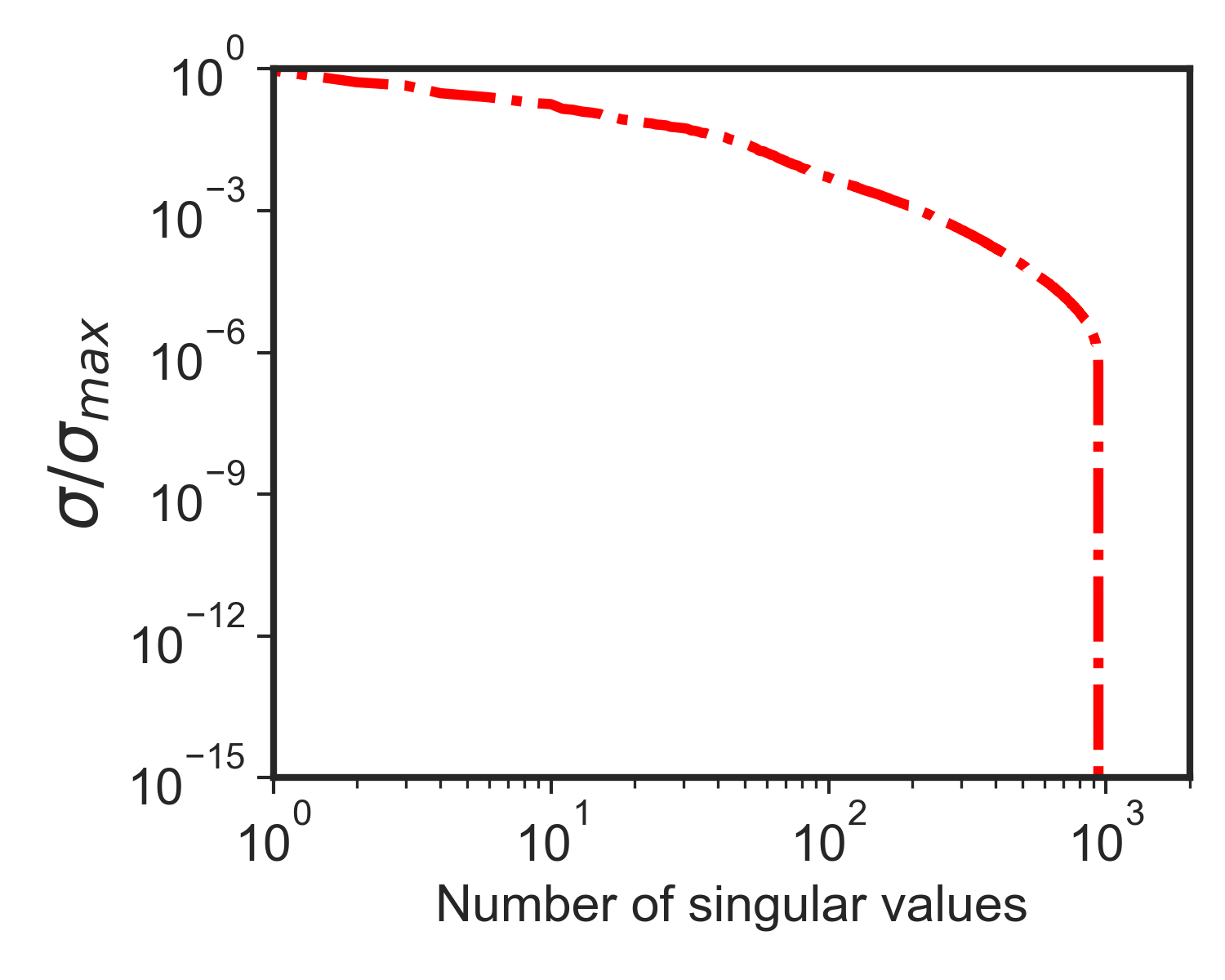}\hfill
\caption{The singular values ($\sigma$) are obtained from a set of $10^{3}$ whitened glitches using SVD \cite{van1996matrix}, normalized by the maximum singular values ($\sigma_{\text{max}})$. The glitches are generated from CT-GAN. The spectrum of singular values is seen to fall sharply, implying only a few singular values (e.g., $\ell = 10$), and corresponding basis vectors are sufficient to represent the glitches. See the Fig.\ref{fig:Glitch-Reconstruct-Error} in which the relative reconstruction error for these glitches has been shown based on $\ell = 1, 5, 10$. For performing SVD based matched filtering for glitch templates, we followed the framework presented in \cite{cannon2010singular}.}
\label{fig:sing-values}
\end{figure}
\begin{figure}[!]
\includegraphics[width=1\columnwidth]{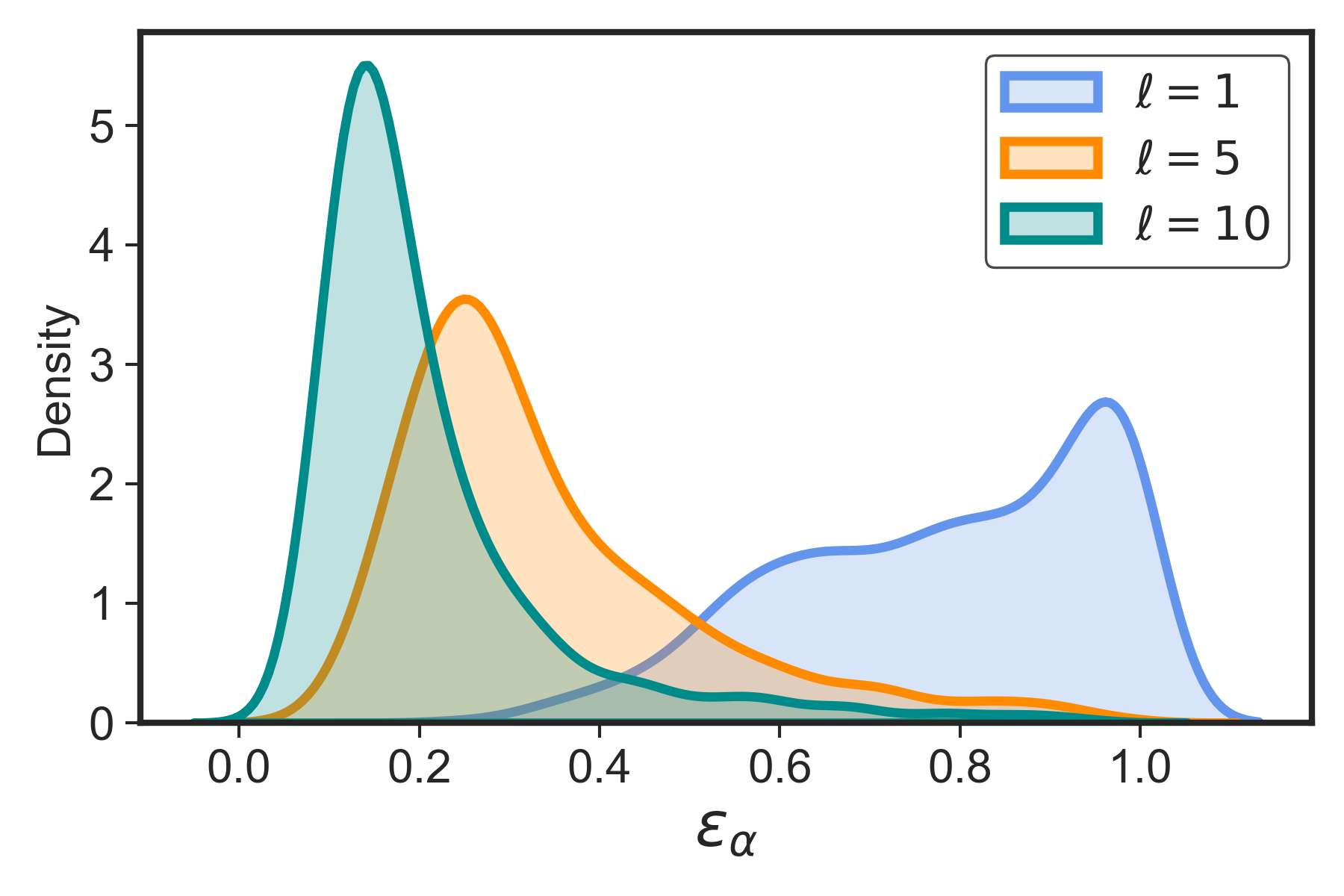}\hfill
\caption{The plot shows the distribution of relative errors for the reconstruction of the $10^{3}$ whitened glitches generated using CT-GAN. The relative error ($\epsilon_{\alpha}$) is calculated for each case.}
\label{fig:Glitch-Reconstruct-Error}
\end{figure}
\begin{figure}[!]
\includegraphics[width=1\columnwidth]{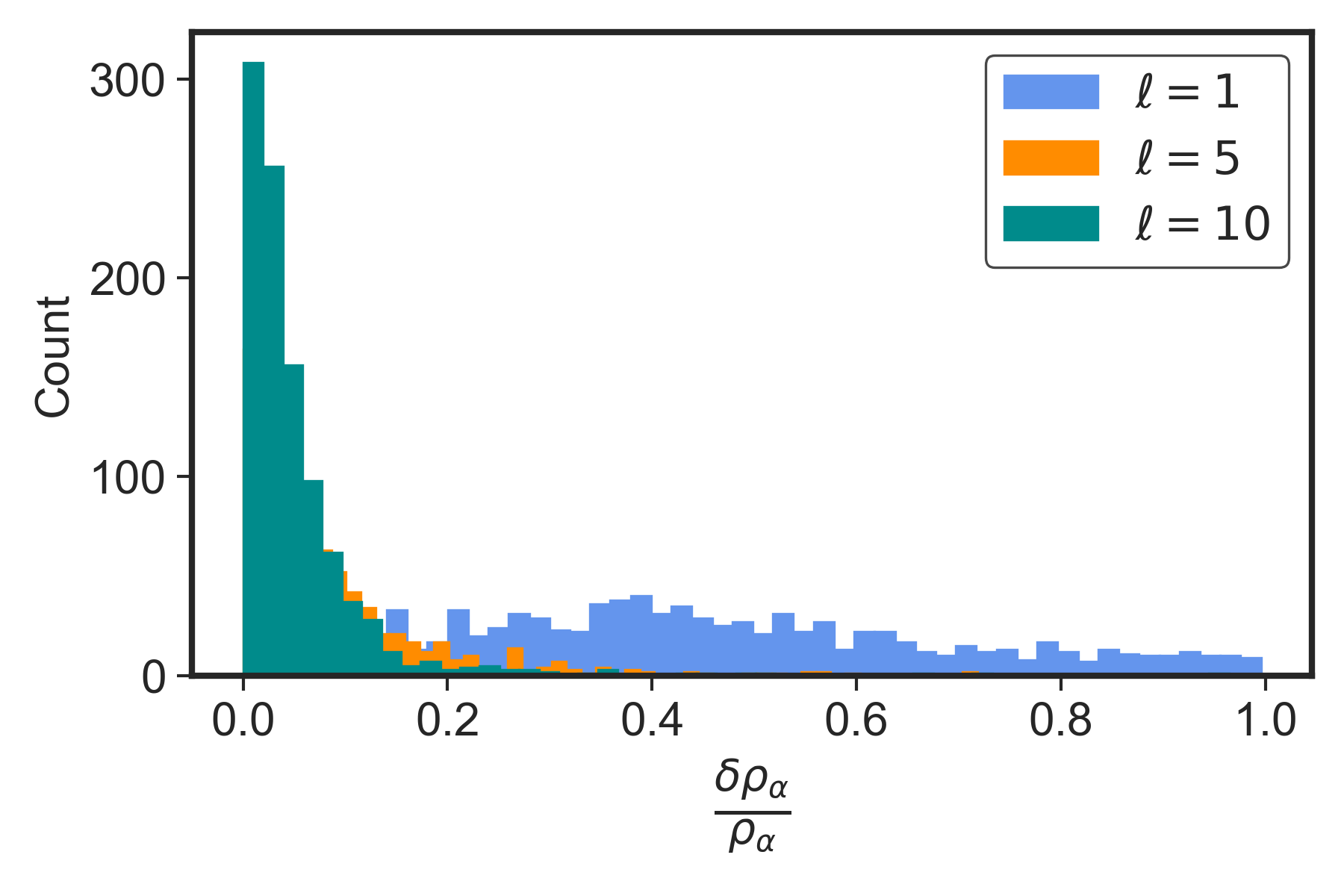}\hfill
\caption{This figure shows the histogram based on the fractional SNR loss ($\frac{\delta \rho_{\alpha}}{\rho_{\alpha}}$) for a set of glitches ($10^{3}$). For each glitch template, the SNR time-series were obtained based on (a) Standard matched-filter scheme and (b) SVD based matched filtering framework presented in \cite{cannon2010singular} by varying the top-basis numbers as $\ell = 1, 5, 10$ respectively. }
\label{fig:SNR-Reconstruct-Error}
\end{figure}
\item \textit{Mock data challenges:} With our methodology, we are able to generate glitches in the time domain. The user could generate as many glitches as necessary, selecting the ones that represent best the real distribution and injecting them in real detector noise to create a realistic data challenge. Moreover, since certain anomalies are generated, those can also be selected to stress-test analysis algorithms.\\
As a preliminary test, we inject some blip glitches in the O3a data to evaluate how they will impact the long-duration analysis with a dedicated neural network called ALBUS \citep{Boudart:2022xib}. For visualization, we present the output in the right panel of Fig.\ref{fig:test_injection}. Since a time resolution is much larger than the glitch duration (i.e., $<0.3$ s), the injected glitch appears as a vertical line. The structure of the glitch is fully recovered and allows to reveal the detection capability of ALBUS.
As suggested in \cite{CGAN_bursts}, when learning different classes of glitches, we could also interpolate between them to generate hybrid classes. This hybrid dataset could be employed to discover unknown classes of glitches and improve the efficiency of detection algorithms.
\begin{figure}[!htb]
\centering
\includegraphics[trim={0.5cm 0.5cm 1.5cm 0.5cm}, clip, scale=0.42]{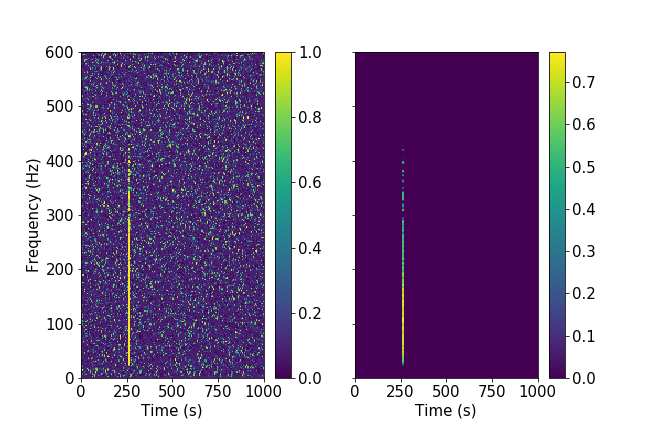}
\caption{Example of glitch injection. The left image shows the input time-frequency map while the right panel shows the output of ALBUS.}
\label{fig:test_injection}
\end{figure}
\item \textit{New glitch detection:} Once the network has learned the underlying distribution of the data, with certain modifications, it can output how likely it is for an unknown signal to belong to the known distribution. This metric can detect anomalous generations and provide feedback to classification algorithms. For example, with this information \textit{Gravity Spy} could re-classify certain anomalies, which could imply the definition of new glitch classes and their further characterization.  
\item \textit{Improving glitch classification:} One of the main challenges of working with real data is to deal with imbalanced data sets. With our methodology, once more classes are learned, we could generate balanced data sets to improve the accuracy of classification algorithms.
\end{enumerate}

\section{Conclusion}\label{sec:conclusion}

In this work, we have developed a methodology to generate artificial blip glitches from real data using a ML algorithm known as GAN. To be able to generate these glitches, the input blips need to be processed: the signals are selected from {Gravity Spy} data to be reconstructed with {BayesWave} and smoothed with the rROF algorithm. Because of this heavy processing, only around $66 \%$ and $50 \%$ of the initial data from L1 and H1 is preserved. 

Due to the instability of GAN algorithms, in this particular research, we trained a CT-GAN \cite{CTGAN}. The network uses Wasserstein distance as a loss function, which allows it to train its discriminator to optimality. The network is penalized heavily to avoid training instabilities and to learn the underlying distribution of blips accurately. 

To assess the performance of CT-GAN, we generate a population of $10^{3}$ blip glitches for both H1 and L1. The quality measurements employed are \textit{Gravity Spy} classifier and similarity distances, namely,  Wasserstein distance ($W_{1}$), match function ($M_{f}$) and normalized cross-covariance ($k$). The results of these metrics indicate that the neural network was able to learn the underlying distribution of blip glitches from H1 and L1, despite the presence of some anomalous generations due to imperfections of the input data set. Furthermore, it has been observed that the similarity distances are able to detect miss-classifications from glitch classifiers.

In this proof-of-concept investigation, we have demonstrated that it is possible to isolate blip glitches from their surrounding noise and learn their underlying distribution with an ML-based method in the time domain, providing several examples of its usage. This methodology allows us to generate better quality data, and it provides us with flexibility that would be challenging to achieve with Q-transforms. The long-term goal of this investigation is to learn other classes of glitches and create an open-source interface for producing real data in the time domain.    

\section*{Acknowledgement}
The authors thank Chris Messenger, Siddharth Soni, Jess McIver, Marco Cavaglia, Alejandro Torres-Forné and Harsh Narola for their useful comments. V.B. is supported by the Gravitational Wave Science (GWAS) grant funded by the French Community of Belgium, and M.L., S.C and A.R are supported by the research program of the Netherlands Organisation for Scientific Research (NWO). The authors are grateful for computational resources provided by the LIGO Laboratory and supported by the National Science Foundation Grants No. PHY-0757058 and No. PHY-0823459. This material is based upon work supported by NSF's LIGO Laboratory which is a major facility fully funded by the National Science Foundation. 
\bibliography{apssamp}
\onecolumngrid
\appendix
\section{Results for blip distribution from L1}
\label{sec:appendix}
\begin{figure}[ht]
\centering
\includegraphics[width=0.5\textwidth]{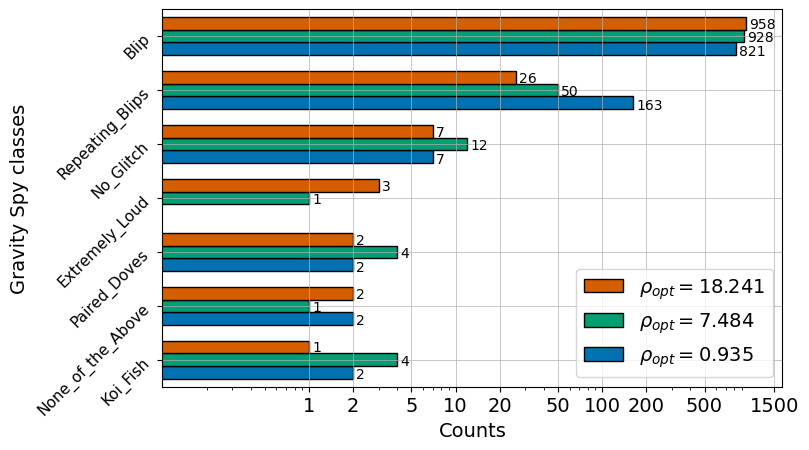}
\caption{Histogram of predicted \textit{Gravity Spy} classes for $10^3$ blips from L1.}
\label{fig:class_gs_l1}
\end{figure}

\begin{figure*}[!]

\subfloat[{Wasserstein distance ($W_1$)}]{%
\includegraphics[width=0.33\textwidth]{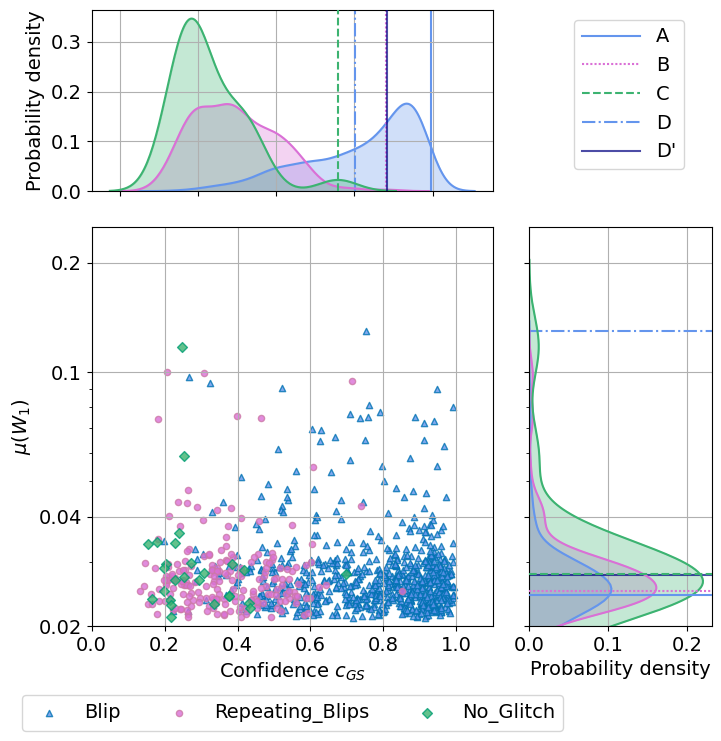}%
}\hfill
\subfloat[{Match fucntion ($M_f$)}]{%
\includegraphics[width=0.33\textwidth]{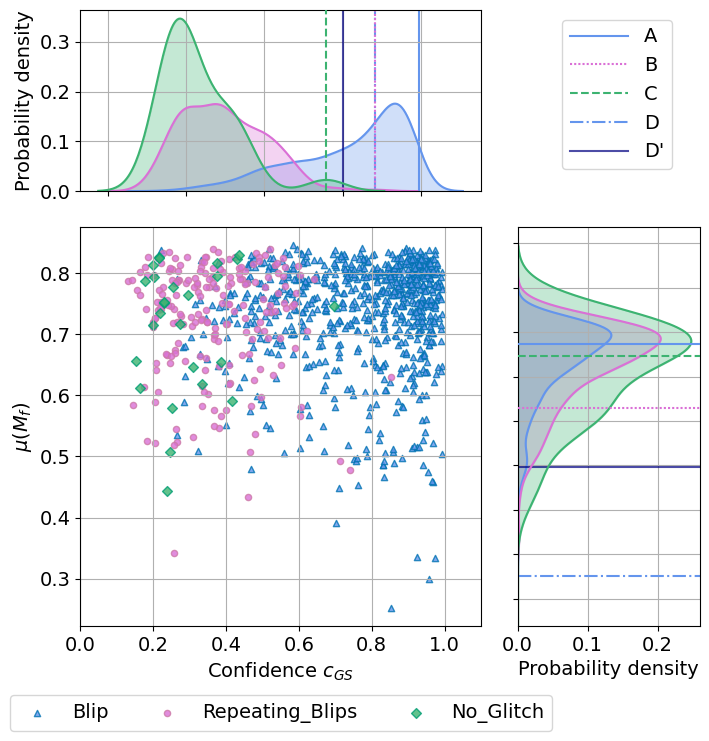}%
}\hfill
\subfloat[{Normalised cross-covariance ($k$)}]{%
\includegraphics[width=0.33\textwidth]{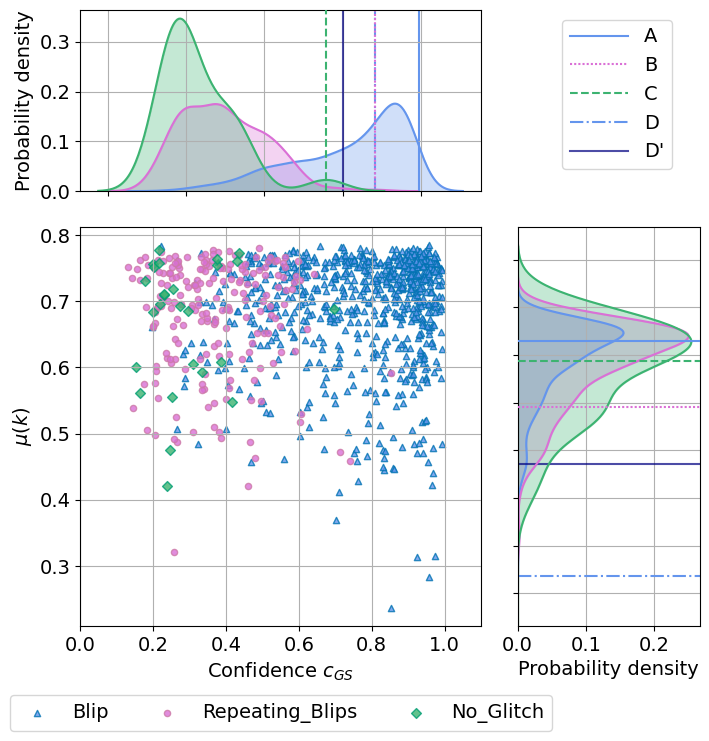}%
}
\caption{Joint and marginal distribution of \textit{Gravity Spy} confidence $c_{GS}$ at $\rho_{opt}=18.46$ against different metrics for different glitch classes for L1: $Blip$, $Repeating\_Blips$ and $No\_Glitch$. We mark in the marginal distributions selected glitches A (solid blue), B (dotted pink), C (dashed green) and D (dash dotted blue).}
\label{fig:gs_metrics_l1}
\end{figure*}

This appendix presents the results of blips from the L1 distribution, which are compatible with the H1 population. In Fig.\ref{fig:class_gs_l1}, we  present a histogram of the classes assigned by \textit{Gravity Spy} to a population of $10^{3}$ artificial blips. As in Section \ref{sec:results}, we can also observe that the three dominant classes are $Blip$, $Repeating\_Blips$ and $No\_Glitch$, and as we increase the optimal SNR $\rho_{opt}$, the number of artificial glitches classified as $Blip$ increases.
As we stated before, \textit{Gravity Spy} classifier seems to be biased towards $Blip$ class, since at very low $\rho_{opt}$, the network will be unable to see the glitches. Another interesting question would be to assess the influence of the detector noise in the classification task of {Gravity Spy}. Similarly to Fig. \ref{fig:gs_metrics}, we present in Fig. \ref{fig:gs_metrics_l1} the confidence of \textit{Gravity Spy} as a function of alternative metrics for the dominant classes. 
In Fig.\ref{fig:gs_metrics_l1}, we can also observe that there is no apparent correlation between the measurements and the confidence provided by \textit{Gravity Spy} classifier. To inspect the results, we select certain glitches according to the definitions in \ref{sec:metrics_definition}. Note that the anomalous glitch found by Wasserstein distance (labeled as D) does not coincide with the one found by match function and normalized cross-covariance (labeled as D'). \textit{Gravity Spy} was able to correctly classify with a high confidence glitch A and B, but glitches C, D, and D' are misclassified. 
\begin{figure*}[htb]
\hfill
\subfloat[{Glitch A}]{%
\includegraphics[width=0.204\columnwidth]{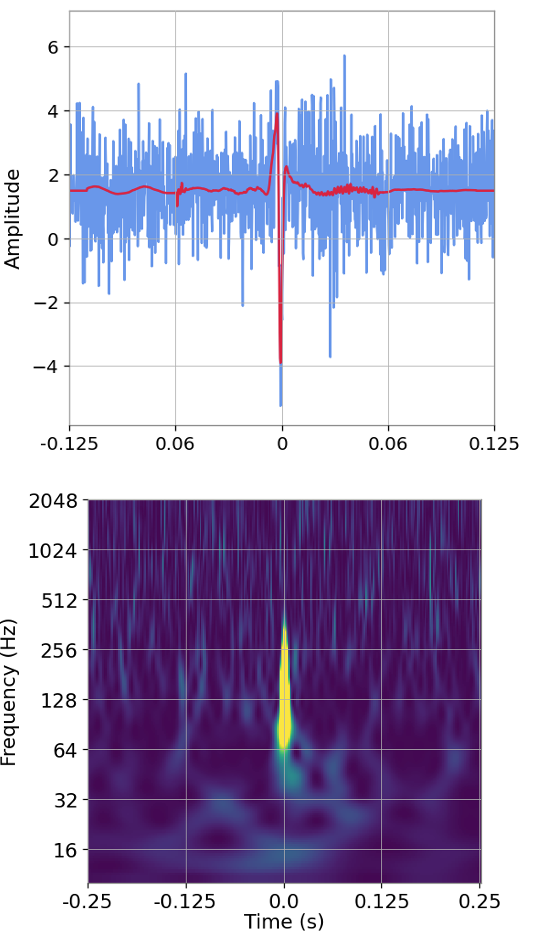}%
}\hfill
\subfloat[{Glitch B}]{%
\includegraphics[width=0.19\columnwidth]{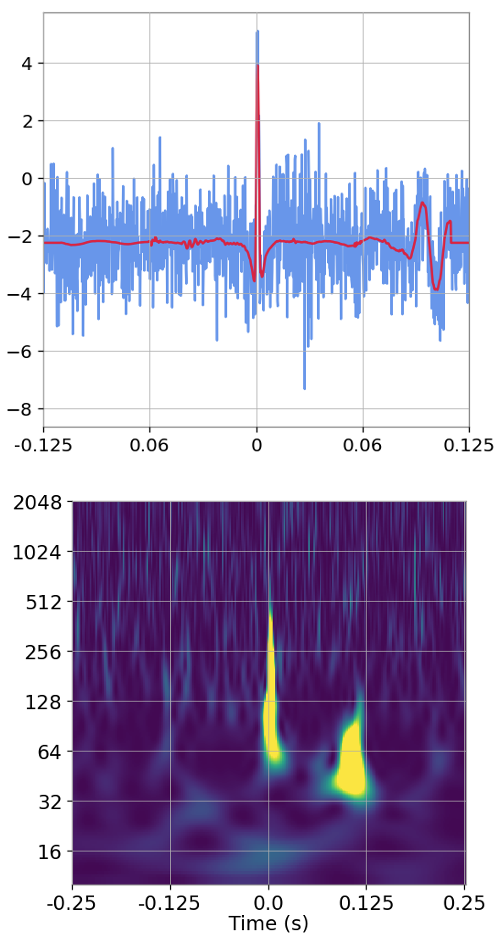}%
}
\subfloat[{Glitch C}]{%
\includegraphics[width=0.19\columnwidth]{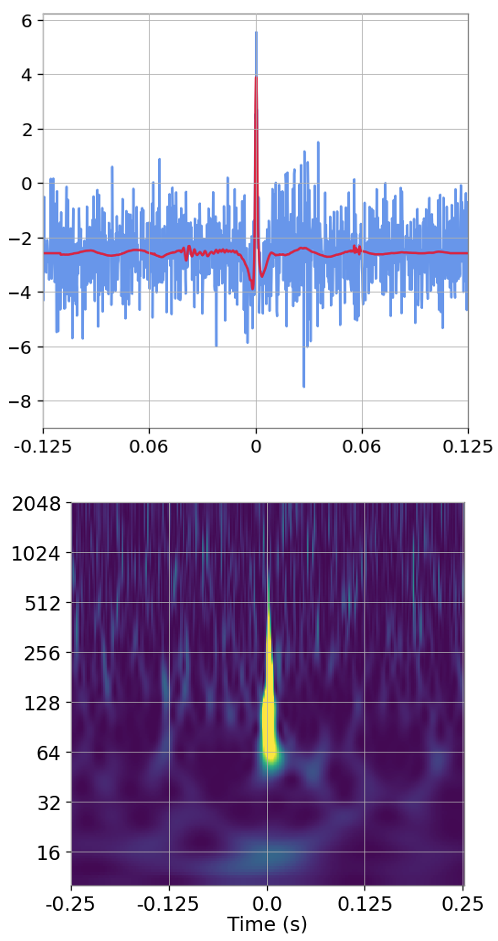}%
}\hfill
\subfloat[{Glitch D}]{%
\includegraphics[width=0.195\columnwidth]{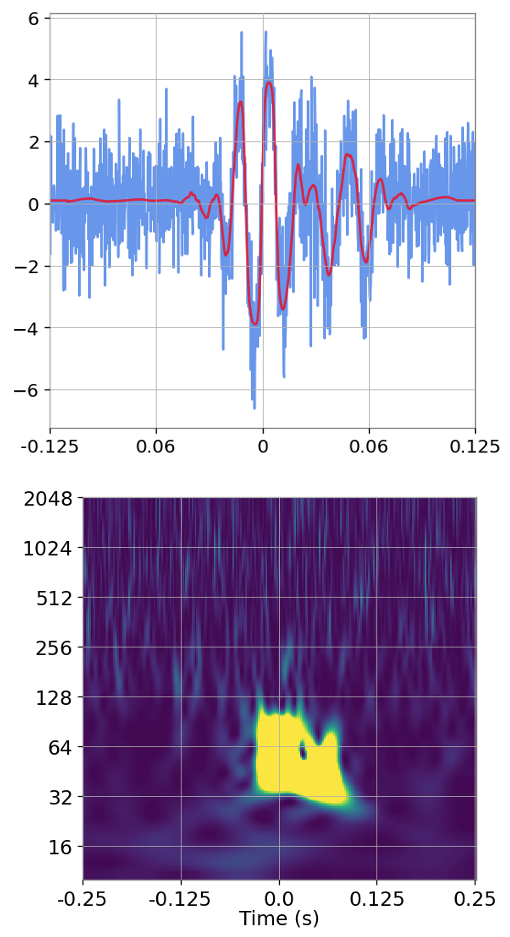}%
}
\subfloat[{Glitch D'}]{%
\includegraphics[width=0.19\columnwidth]{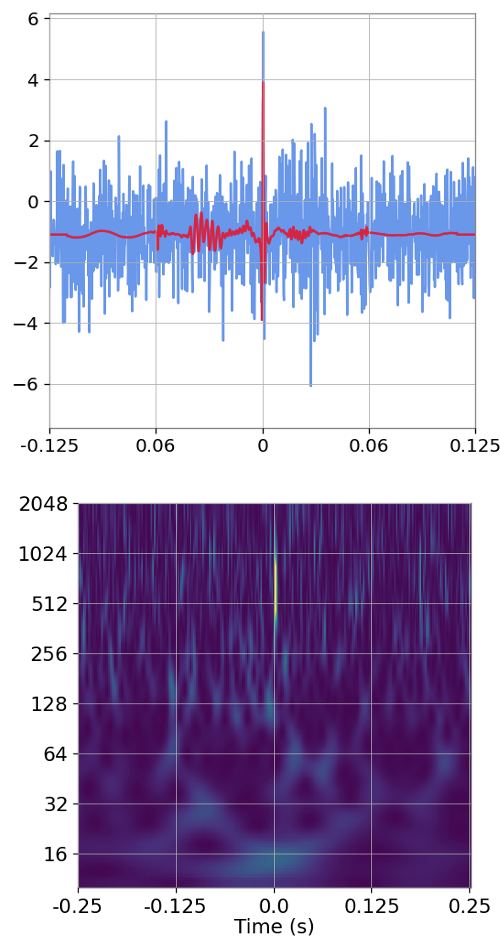}%
}\hfill
\caption{Time series representation (top row) and Q-scan representation of selected glitches from L1}
\label{fig:generations_l1}
\end{figure*}
For visualization and a better understanding of the results, we plot in Fig.\ref{fig:generations_l1} the Q-transforms and the time series injected in real whitened noise of the selected glitches. While glitch A is classified by \textit{Gravity Spy} as a perfect glitch, glitch C is miss-classified as $No\_Glitch$, although their Q-transforms look similar. It is interesting to mention that the GAN was able to generate a $Repeating\_Blip$ because some repeating blips are present in the input data set.  

Glitches D and D', which are misclassified by \textit{Gravity Spy}, are situated in the tail of the distribution of the similarity distances. While glitch D has a shape very different from a standard blip, glitch D' has a very narrow peak.

\end{document}